\documentclass[12pt]{article}

\usepackage[totalwidth=460truept,totalheight=585truept]{geometry}
\usepackage{latexsym,graphicx,amsfonts,amssymb,amsmath}
\usepackage[hypertexnames=false,hidelinks]{hyperref}

\global\arraycolsep=1truept

\begin{document}

\null

\vskip1truecm

\begin{center}
{\huge \textbf{Diagrammar of Physical}}

\vskip.8truecm

{\huge \textbf{\ and Fake Particles}}

\vskip.8truecm

{\huge \textbf{and Spectral Optical Theorem}}

\vskip1truecm

\textsl{Damiano Anselmi}

\vskip .1truecm

\textit{Dipartimento di Fisica \textquotedblleft Enrico Fermi", Universit%
\`{a} di Pisa}

\textit{Largo B. Pontecorvo 3, 56127 Pisa, Italy}

\textit{and INFN, Sezione di Pisa,}

\textit{Largo B. Pontecorvo 3, 56127 Pisa, Italy}

damiano.anselmi@unipi.it

\vskip1truecm

\textbf{Abstract}
\end{center}

We prove spectral optical identities in quantum field theories of physical
particles (defined by the Feynman $i\epsilon $ prescription) and purely
virtual particles (defined by the fakeon prescription). The identities are
derived by means of purely algebraic operations and hold for every
(multi)threshold separately and for arbitrary frequencies. Their major
significance is that they offer a deeper understanding on the problem of
unitarity in quantum field theory. In particular, they apply to
\textquotedblleft skeleton\textquotedblright\ diagrams, before integrating
on the space components of the loop momenta and the phase spaces. In turn,
the skeleton diagrams obey a spectral optical theorem, which gives the usual
optical theorem for amplitudes, once the integrals on the space components
of the loop momenta and the phase spaces are restored. The fakeon
prescription/projection is implemented by dropping the thresholds that
involve fakeon frequencies. We give examples at one loop (bubble, triangle,
box, pentagon and hexagon), two loops (triangle with \textquotedblleft
diagonal\textquotedblright , box with diagonal) and arbitrarily many loops.
We also derive formulas for the loop integrals with fakeons and relate them
to the known formulas for the loop integrals with physical particles.

\vfill\eject

\section{Introduction}

\label{intro}\setcounter{equation}{0}

Unitarity is a key requirement to claim that a quantum field theory has
chances to be fundamental, together with locality and renormalizability.
Generically, locality and renormalizability can be phrased in simple terms:
the Lagrangian should be polynomial in the fields and their derivatives and
obey the power counting criterion. Moreover, the theorem of locality of
counterterms shows that there is a nontrivial connection between
renormalizability and locality. Unitarity, on the other hand, still lacks an
understanding in equally simple terms. In this paper we overcome this
drawback and extend the classical results \cite%
{cutkosky,veltman,thooft,diagrammar,diagrammatica} in several directions.

We show that unitarity in quantum field theory can be reduced to a set of
algebraic identities, which do not require to integrate on the space
components of the loop momenta, or the phase spaces in cut diagrams, and
hold for each physical threshold separately. The key ingredient is a proper
threshold decomposition, because different thresholds do not talk to one
another.

The gain in insight is important. Among the other things, we can identify
the purely virtual (off-shell) content of amplitudes and diagrams. We can
switch from the Feynman prescription to the fakeon prescription \cite%
{LWgrav,fakeons} by simply dropping all the thresholds that involve one or
more frequencies associated with the legs that we want to quantize as
fakeons (that is to say, purely virtual particles). We can easily generalize
the identities to propagators with arbitrary residues and arbitrary
frequencies, as well as vertices with arbitrary derivative structures (as
long as the classical Lagrangian stays Hermitian). We can show that the
spectral identities hold for thick fakeons as well (fakeons with finite
\textquotedblleft widths\textquotedblright\ at the tree level\footnote{%
For a review, see \cite{causalityQG}.}), which are typically originated by
higher-derivative Lagrangians and can be used to reformulate the Lee-Wick
models \cite{leewick,lee,nakanishi,CLOP,grinstein} as models of particles
and fakeons \cite{LWformulation,LWunitarity}. We can even introduce certain
details of the experimental apparatus without violating the spectral optical
identities. For example, we can prove unitarity in the presence of fakeons
when the energy resolution of detectors is taken into account, as a cutoff
for the infrared divergences of massless fields \cite{infred}. It is also
possible to include the energy resolution of fakeons in a unitary way.

Purely virtual particles \cite{wheelerons}, or \textquotedblleft
fakeons\textquotedblright , can be used to formulate a consistent theory of
quantum gravity \cite{LWgrav}, which is experimentally testable thanks to
its predictions in inflationary cosmology \cite{ABP}. More generally, they
can be used to solve the problem of ghosts in higher-derivatives theories.
However, they can also be employed in collider physics \cite{Tallinn1} to
evade common constraints and offer new ways to solve discrepancies with
data, as in the problem of the muon anomalous magnetic moment \cite{Tallinn2}%
.

\bigskip

Everything we do in this paper involves nothing more than purely algebraic
operations. Despite some unavoidably lengthy expressions, the identities we
write can be checked straightforwardly. First, we integrate on the loop
energies. This is the only integral we need to do. We can view it as an
algebraic operation as well, thanks to the residue theorem. Then, we ignore
the integrals on the space components of the loop momenta and the phase
spaces, i.e., work on the \textquotedblleft skeleton\textquotedblright\ of
the diagram. Third, we eliminate the pseudothresholds, which are unphysical,
because they involve differences of frequencies. Once we remain with the
physical thresholds, we proceed with the threshold decomposition, which
provides a separate optical identity for each threshold. Summing the
identities associated with a loop diagram, we derive the spectral optical
theorem obeyed by its skeleton. Integrating the spectral optical theorem on
the space components of the loop momenta and the phase spaces, we obtain the
usual optical theorem for amplitudes, which proves unitarity.

Besides proving the identities and the spectral optical theorem for
arbitrary diagrams, we give a large number of examples that show how the
threshold decomposition works. Apart from the bubble and the triangle, which
are relatively straightforward, we study the box, the pentagon and the
hexagon, at one loop. At two loops we study the triangle with
\textquotedblleft diagonal\textquotedblright\ and the box with diagonal.
These examples cover the needs of most calculations in high-energy physics
phenomenology. Whole classes of diagrams with arbitrarily many loops are
included straightforwardly. The formulas we obtain for the loop integrals
with fakeons and physical particles can be related to the formulas of
ordinary loop integrals (defined by the Feynman prescription everywhere) and
implemented in softwares like FeynCalc, FormCalc, LoopTools and Package-X 
\cite{calc}.

The paper is organized as follows. In section \ref{strategy} we collect the
basic definitions and the strategy of the calculations. In sections \ref%
{bubbles}, \ref{triangleT} and \ref{boxG} we study the bubble diagram (and
its multibubble versions), the triangle diagram and the box diagram,
respectively. In section \ref{multi} we extend the results to classes of
multiloop diagrams. In section \ref{boxwdiag} we study the box diagram with
diagonal. In section \ref{smartsec} we provide further insight into the
algebraic structure of the spectral optical identities and derive the
formulas of the pentagon and the hexagon. In section \ref{arbres} we
generalize the results to propagators with arbitrary residues. In section %
\ref{degenerate} we extend the proof to diagrams with derivative vertices
(integrands with nontrivial numerators) and degenerate diagrams (diagrams
with powers of identical propagators or subdiagrams). In section \ref{thick}
we generalize the proofs to propagators with thick fakeons. In section \ref%
{massless} we show how to treat massless fields and their infrared
divergences. Section \ref{conclusions} contains the conclusions.

\section{Definitions and strategy}

\label{strategy}\setcounter{equation}{0}

In this section we collect the basic definitions we need and the strategy of
the calculations. Since the matter is technically involved, not all the
subtleties we anticipate here can be appreciated right away. Nevertheless,
it is convenient to collect them in a dedicated place to allow a quick back
and forth, for a better understanding of the next sections.

Unitarity is the statement that the scattering matrix ($S$ matrix) is
unitary. Writing $S=1+iT$, where $T$ is the transition amplitude (and $iT$
is the sum of loop diagrams), we can write the unitarity equation $S^{\dag
}S=1$ in the equivalent form%
\begin{equation}
2\hspace{0.01in}\mathrm{Im}T=T^{\dag }T,  \label{optical}
\end{equation}%
which is known as optical theorem. Proving unitarity, or the optical
theorem, is crucial to establish that a quantum field theory is consistent
as a fundamental theory.

More powerful statements can be proved, in general, such as diagrammatic
versions of (\ref{optical}), which hold for each loop integral separately.
They can be expressed as relations%
\begin{equation}
G+\bar{G}+\sum_{c}G_{c}=0  \label{diagroptical}
\end{equation}%
among a given diagram $G$, its complex conjugate $\bar{G}$ and certain
\textquotedblleft cut diagrams\textquotedblright\ $G_{c}$. The cut diagrams
are diagrams divided into two portions by a cut passing through a number of
propagators (which become \textquotedblleft cut
propagators\textquotedblright ). One portion stands of the $T$ of (\ref%
{optical}) and the other one (which is normally shadowed) stands for $%
T^{\dag }$. Like $G$ and $\bar{G}$, the cut diagrams can be defined by means
of diagrammatic rules, which, in addition to the ordinary propagators and
vertices, involve cut propagators, complex conjugate propagators and complex
conjugate vertices. The rules are given in subsection \ref{diagra}.

In ref. \cite{ACE} it was shown that the diagrammatic version (\ref%
{diagroptical}) of the optical theorem can be derived algebraically and
expressed by means of algebraic equations, which do not require to integrate
on the space components of the loop momenta or phases spaces\footnote{%
Some integrals on the loop space momenta of $G$ and $\bar{G}$ appear as
integrals on phase spaces in $G_{c}$. They are those interested by the cut
propagators, which put the particles on shell. They stand for the
contraction between $T^{\dag }$ and $T$ of formula (\ref{optical}).}. In
this paper we exploit these properties to the fullest.

\begin{figure}[t]
\begin{center}
\includegraphics[width=14truecm]{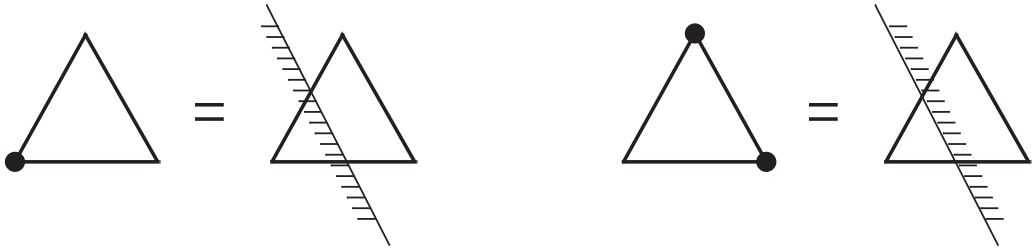}
\end{center}
\caption{Marked diagrams and cut diagrams. The shadowed portion stands for $%
T^{\dagger }$, the unshadowed one for $T$.}
\label{cutcut}
\end{figure}

\begin{figure}[t]
\begin{center}
\includegraphics[width=14truecm]{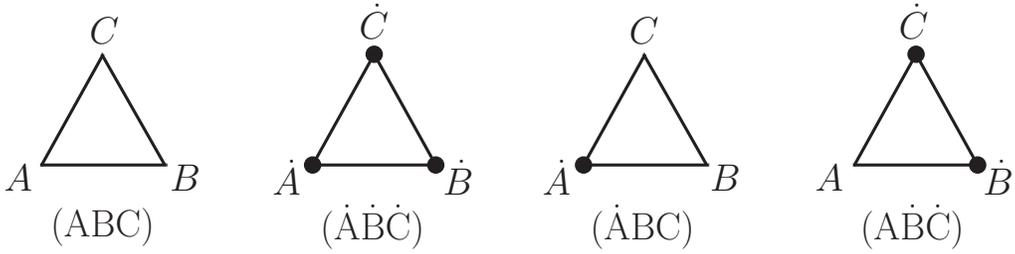}
\end{center}
\caption{Triangle uncut and cut diagrams (in marked notation)}
\label{cuttitutti}
\end{figure}

As in \cite{ACE}, it is convenient to denote the cut diagrams in
\textquotedblleft dual\textquotedblright\ form, by means of marked and
unmarked vertices. In fig. \ref{cutcut} we show the relation between marked
diagrams and cut diagrams. In fig. \ref{cuttitutti} we show the marked
versions of the triangle diagram, up to permutations of the vertices. 

This convention allows us to identify the diagrams by means of simple
\textquotedblleft words\textquotedblright . Let A, B, C, etc.,\ label the
unmarked vertices and \.{A}, \.{B}, \.{C}, etc., the marked vertices.\ The
propagators are segments. Segments like AB, BC, etc., are ordinary
propagators, while segments like \.{A}\.{B}, \.{B}\.{C}, etc., are complex
conjugate propagators. Segments like \.{A}B, B\.{C}, etc., are cut
propagators with positive energies flowing from the vertices without the dot
to the vertices with the dot.

The loops are enclosed between parentheses. A one-loop (cut or uncut)
diagram is a \textquotedblleft word\textquotedblright\ between parentheses,
built with letters. For example: (AB) is the uncut bubble diagram $\rangle 
\hspace{-0.18em}{\bigcirc \hspace{-0.18em}\langle }$, where A denotes the
left vertex and B is the right vertex. Then, (\.{A}\.{B}) denotes the
complex conjugate diagram, while (\.{A}B) denotes the cut diagram with
positive energies flowing from B to A and (A\.{B}) is the cut diagram with
positive energies flowing from A to B.

Similarly, (ABC) is the uncut triangle diagram, (\.{A}\.{B}\.{C}) is its
complex conjugate, (\.{A}BC) is the cut diagram with positive energies
flowing to A, etc.

Diagrams with more loops are denoted by means of products of one-loop words
in parentheses, repeating the vertices that belong to more loops. For
example, the double bubble $\rangle \hspace{-0.18em}{\bigcirc \hspace{-0.18em%
}{|\hspace{-0.18em}{\bigcirc \hspace{-0.18em}\langle }}}$ is (AB)(BC), its
conjugate is (\.{A}\.{B})(\.{B}\.{C}), its cut versions are (\.{A}B)(BC), (A%
\.{B})(\.{B}C), etc.

The legs are labeled by numbers: the box diagram, which is concisely denoted
by (ABCD), can be denoted by (A1B2C3D4) in extended form, where 1 is the leg
AB, 2 is the leg BC, 3 is the leg CD and 4 is the leg DA. Similarly,
(AB)(BC) may stand for (A1B2)(B3C4), etc. Different \textquotedblleft
words\textquotedblright\ may correspond to the same diagram, but this is not
going to cause trouble.

The diagrammatic optical theorem (\ref{diagroptical}) is the statement that,
given a diagram $G$, the sum of all its marked versions (including $G$
itself) vanishes. For example, in the case of the triangle the sum of the
diagrams shown in fig. \ref{cuttitutti} plus the cyclic permutations of the
last two is equal to zero. Here we generalize the theorem to spectral
optical identities, which separately hold for each threshold. To achieve
this result, we need to work out a neat threshold decomposition of the
\textquotedblleft skeleton\textquotedblright\ diagrams, defined below.

In the first part of our investigation, we concentrate on the loop integrals
defined by the Feynman $i\epsilon $ prescription everywhere. As soon as
their threshold decomposition is worked out, it is relatively
straightforward to apply the fakeon prescription/projection to it.

We do not need to handle the ultraviolet divergences, but, for the sake of
precision, we understand that the loop integrals are defined by means of the
dimensional regularization \cite{dimreg} as follows: the integrals on the
loop energies are \textit{defined} by the residue theorem, while the
integrals on the space components of the loop momenta are defined as the
dimensionally regularized ones in $D-1$ continued dimensions. 

\subsection{Skeletons}

The only integrals we compute in this paper are those on the loop energies.
Their evaluation amounts to an algebraic operation, thanks to the residue
theorem. If we drop the integrals over the space components of the loop
momenta (and the phase spaces, in cut diagrams), as well as certain overall
factors, we obtain the \textit{skeletons}.

If $G(p)$ denotes an ordinary loop integral (with all the propagators
defined by the Feynman $i\epsilon $ prescription), the skeleton $G^{s}$ of $%
G $ is defined by the formula 
\begin{equation}
G(p)=\int \prod\limits_{l=1}^{L}\frac{\mathrm{d}^{D-1}\mathbf{k}_{l}}{(2\pi
)^{D-1}}\left( \prod\limits_{a=1}^{I}\frac{1}{2\omega _{a}}\right) G^{s}(p,%
\mathbf{k}),
\end{equation}%
where $p_{a}^{\mu }=(e_{a},\mathbf{p}_{a})$ are the external momenta (one
for each \textit{internal} leg), $k_{l}^{\mu }=(k_{l}^{0},\mathbf{k}_{l})$
are the loop momenta, $L$ is the number of loops, $I$ is the number of
internal legs and $\omega _{a}$ is the frequency associated with the $a$th
internal leg.

For example, the basic one-loop integrals 
\begin{equation}
G_{N}=\int \frac{\mathrm{d}^{D}k}{(2\pi )^{D}}\prod\limits_{a=1}^{N}\frac{1}{%
(k-p_{a})^{2}-m_{a}^{2}+i\epsilon _{a}},
\end{equation}%
define the one-loop skeletons%
\begin{equation}
G_{N}^{s}=\int \frac{\mathrm{d}k^{0}}{2\pi }\prod\limits_{a=1}^{N}\frac{%
2\omega _{a}}{(k-p_{a})^{2}-m_{a}^{2}+i\epsilon _{a}}=\int \frac{\mathrm{d}%
k^{0}}{2\pi }\prod\limits_{a=1}^{N}\frac{2\omega _{a}}{(k^{0}-e_{a})^{2}-%
\omega _{a}^{2}+i\epsilon _{a}},  \label{GsN}
\end{equation}%
so that%
\begin{equation}
G_{N}=\int \frac{\mathrm{d}^{D-1}\mathbf{k}}{(2\pi )^{D-1}}\left(
\prod\limits_{a=1}^{N}\frac{1}{2\omega _{a}}\right) G_{N}^{s},
\end{equation}%
where the frequencies are $\omega _{a}=\sqrt{(\mathbf{k}-\mathbf{p}%
_{a})^{2}+m_{a}^{2}}$.

Note that $L$ momenta $p_{a}$ are redundant, since they can be eliminated by
translating the loop momenta. The redundant notation allows us to treat all
the internal legs equally, which is more convenient for the derivations of
this paper.

Given complexified external momenta $p$, if the skeleton $G^{s}(p,\mathbf{k}%
) $ is regular for every values of the space components $\mathbf{k}_{l}$ of
the loop momenta within their integration domains, then $G(p)$ is analytic
in $p$. The most important analytic region is the Euclidean one, where the
energies are imaginary.

The $p$ domains where $G(p)$ is not analytic are found by studying the
singularities of $G^{s}(p,\mathbf{k})$. Important domains of non analyticity
are identified by inequalities of the form%
\begin{equation}
P_{E}^{2}\geqslant \left( \sum_{a\in J}m_{a}\right) ^{2},  \label{thresh}
\end{equation}%
where $P_{E}$ is a linear combination of external momenta and $J$ is a
subset of internal legs. The right-hand side of this formula is an optical
threshold, namely the minimum invariant mass of a physical process where the
external particles involved in $P_{E}$ produce the internal particles of the
subset $J$, turning them from virtual to real\footnote{%
A fakeon is purely virtual, i.e., it can never become real. If a leg of the
set $J$ is defined as a fakeon (see below for details), the threshold
appearing on the right-hand side of (\ref{thresh}) is not associated with a
physical process. It is associated with the mathematical violation of
analyticity and other physical properties, typical of fakeons, which we do
not need to detail here. We may call it a \textquotedblleft fake
threshold\textquotedblright\ and say that it is associated with a fake
physical process.}. These thresholds are called \textquotedblleft
optical\textquotedblright , because they contribute to the optical theorem.
Other physical thresholds exist, which do not participate in the optical
theorem. They are called \textit{anomalous} thresholds \cite{anomalousthre}.

After integrating on the loop energies, the skeletons are sums of terms that
involve denominators of the forms%
\begin{equation}
\tilde{D}_{\text{phys}}=E+\sum_{a\in J}\tilde{\omega}_{a},\qquad \tilde{D}_{%
\text{pseudo}}=E+\sum_{a\in J}\tilde{\omega}_{a}-\sum_{b\in J^{\prime }}%
\tilde{\omega}_{b},  \label{Dtile}
\end{equation}%
where $\tilde{\omega}_{a}=\omega _{a}-i\epsilon _{a}$, $E$ is a linear
combination of the external energies $e_{c}$ and $J$, $J^{\prime }$ are
nonempty subsets of internal legs. Defining%
\begin{equation}
D_{\text{phys}}=\left. \tilde{D}_{\text{phys}}\right\vert _{\epsilon
_{a}\rightarrow 0},\qquad D_{\text{pseudo}}=\left. \tilde{D}_{\text{pseudo}%
}\right\vert _{\epsilon _{a}\rightarrow 0},  \label{Dphys}
\end{equation}%
the zeros of $D_{\text{phys}}$ are associated with the physical thresholds,
while the zeros of $D_{\text{pseudo}}$ are associated with the so-called
pseudothresholds. The $i\epsilon $ prescription on the loop energies picks
the right residues, which ultimately make the pseudothresholds cancel out%
\footnote{%
The physical reason why the denominators $D_{\text{pseudo}}$ must disappear
is that the differences of frequencies contained in them would lead to
processes plagued by instabilities.}. To have control on the
pseudothresholds in the calculations, we must choose the infinitesimal
widths $\epsilon _{a}$ such that all the denominators $\tilde{D}_{\text{%
pseudo}}$ have nonvanishing imaginary parts (otherwise they are not well
prescribed).

The linear combinations of energies and frequencies generated by the
diagrammatics have coefficients 0 and $\pm 1$. This is guaranteed by the
fact that an energy cannot flow into the same line twice (assuming a
\textquotedblleft minimal\textquotedblright\ parametrization of the momenta,
such that every loop momentum $k_{l}$ appears in some leg $a$ with momentum $%
p_{a}-k_{l}$). All the manipulations we make in this paper involve linear
combinations of the type just mentioned.

Moreover, the operations we make on skeletons are manifestly Lorentz
invariant once we resume the integrals over the space components of the loop
momenta and the phase spaces. The thresholds (\ref{thresh}) emerge after
such integrals. Since we work on skeletons here, from now we understand that
the words \textquotedblleft singularity\textquotedblright ,
\textquotedblleft threshold\textquotedblright\ and \textquotedblleft
pseudothreshold\textquotedblright\ may\ also refer to the skeletons.

Not all types of $G^{s}(p,\mathbf{k})$ singularities can be generated, but
only \textquotedblleft diagrammatic\textquotedblright\ ones. As said, the
potential violations of analyticity occur when one or more skeleton
denominators vanish. Since the $G^{s}(p,\mathbf{k})$ denominators are
originated by the diagrammatics, it must be possible to express the
singularities diagrammatically. This means that, if we restore the integrals
on the loop energies, the loop energies circulate in the diagram in the way
prescribed by the diagrammatics. In some manipulations, we may create
contributions with non diagrammatic singularities, which cancel out in the
total. We give an example of a non diagrammatic singularity in formula (\ref%
{nondiaga}). We illustrate the cancellation of the non diagrammatic
singularities when we investigate the box (section \ref{boxG}) and the box
with diagonal (section \ref{boxwdiag}).

The spectral optical theorem we are going to prove reads%
\begin{equation}
G^{s}+\bar{G}^{s}+\sum_{c}G_{c}^{s}=0,  \label{spectropt}
\end{equation}%
where $G^{s}$, $\bar{G}^{s}$ and $G_{c}^{s}$ are the skeletons associated
with $G$, $\bar{G}$ and $G_{c}$, respectively. The spectral optical
identities are the threshold contributions to (\ref{spectropt}), which
vanish separately.

\subsection{Diagrammatic rules}

\label{diagra}

It is convenient to think of each propagator as the sum of two separate
contributions, one for each pole:%
\begin{equation}
\frac{i}{p^{2}-m^{2}+i\epsilon }\rightarrow \frac{i}{2\omega }\left( \frac{1%
}{e-\omega +i\epsilon }-\frac{1}{e+\omega -i\epsilon }\right) ,
\label{propade}
\end{equation}%
where $p^{\mu }=(e,\mathbf{p})$ and $\omega =\sqrt{\mathbf{p}^{2}+m^{2}}$.
The arrow means that we are ignoring contributions that disappear when $%
\epsilon $ tends to zero. If we write all the propagators as shown on the
right-hand side of (\ref{propade}), we can expand the integrand of a diagram
as a sum of terms where each internal leg is associated with a unique pole.
This is very convenient for the analysis of this paper.

Since we work with skeleton diagrams, without integrating over the space
components of the loop momenta or phase spaces, we are free to multiply the
propagators by $\omega $-dependent overall factors. We choose such factors
equal to $2\omega $, to match the definition of skeletons given above. The
propagators and cut propagators of physical particles are then%
\begin{eqnarray}
\phantom{\bullet}%
\raisebox{-2pt}{\resizebox{2cm}{!}{$\overset{p\rightarrow
}{\!{\raisebox{2.5pt}{\rule{33pt}{.6pt}}}}$}}\!{\phantom{\bullet}} &=&\frac{i%
}{e-\omega +i\epsilon }-\frac{i}{e+\omega -i\epsilon },  \notag \\
\bullet 
\raisebox{-2pt}{\resizebox{2cm}{!}{$\overset{p\rightarrow }{
\!{\raisebox{2.5pt}{\rule{33pt}{.6pt}}}}$}}\!{\bullet } &=&\frac{i}{e+\omega
+i\epsilon }-\frac{i}{e-\omega -i\epsilon },  \notag \\
\bullet 
\raisebox{-2pt}{\resizebox{2cm}{!}{$\overset{p\rightarrow }{
\!{\raisebox{2.5pt}{\rule{33pt}{.6pt}}}}$}}\!{\phantom{\bullet}} &=&(2\pi
)\delta (e+\omega ),\qquad \bullet \!%
\raisebox{-2pt}{\resizebox{2cm}{!}{$\overset{\leftarrow
p}{{\raisebox{2.5pt}{\rule{33pt}{.6pt}}}}$}}\!{\phantom{\bullet}}=(2\pi
)\delta (e-\omega ).  \label{propag}
\end{eqnarray}

For definiteness, we assume that each unmarked vertex is equal to $-i$.
Then, each marked vertex is equal to $i$. In section \ref{degenerate} we
show how to extend the results to derivative vertices.

We consider connected, non necessarily one-particle irreducible (1PI)
diagrams $G$. If $G$ is disconnected by cutting an internal line $\hat{I}$,
we say that $\hat{I}$ is a \textquotedblleft link\textquotedblright . Inside
cut diagrams $G_{c}$ we apply this definition to both sides of the cut
separately. Thus, if one side of the cut is disconnected when a certain leg
is broken, we call that leg a link.

Now we give the basic diagrammatic rules for the internal legs that we want
to define as fakeons (denoted by \textquotedblleft f\textquotedblright ).
The cut propagators must vanish:%
\begin{equation}
\bullet 
\raisebox{-2pt}{\resizebox{2cm}{!}{$\overset{p\rightarrow }{
\!\underset{\text{f}}{{\raisebox{2.5pt}{\rule{33pt}{.6pt}}}}}$}}=0,\qquad
\bullet 
\raisebox{-2pt}{\resizebox{2cm}{!}{$\overset{\leftarrow p}{
\!\underset{\text{f}}{{\raisebox{2.5pt}{\rule{33pt}{.6pt}}}}}$}}=0.
\label{cutva}
\end{equation}%
The propagator of a link fakeon leg is just the Cauchy principal value $%
\mathcal{P}$ (which is the classical limit of fakeon prescription \cite%
{causalityQG}): 
\begin{eqnarray}
\text{(link)} &\text{:}&\qquad \phantom{\bullet}%
\raisebox{-2pt}{\resizebox{2cm}{!}{$\overset{p\rightarrow
}{\underset{\text{f}}{{\raisebox{2.5pt}{\rule{33pt}{.6pt}}}}}$}}\!%
\phantom{{\bullet}}=\mathcal{P}\left( \frac{i}{e-\omega }-\frac{i}{e+\omega }%
\right) ,  \notag \\
\text{(link)} &\text{:}&\qquad \bullet 
\raisebox{-2pt}{\resizebox{2cm}{!}{$\overset{p\rightarrow }{
\!\underset{\text{f}}{{\raisebox{2.5pt}{\rule{33pt}{.6pt}}}}}$}}\!{\bullet }=%
\mathcal{P}\left( \frac{i}{e+\omega }-\frac{i}{e-\omega }\right) .
\label{link}
\end{eqnarray}

As for the other fakeon legs, the fakeon prescription will be implemented
after deriving the spectral optical identities of $G^{s}$, by dropping the
thresholds that involve frequencies $\omega $ associated with fakeon legs.

Note that it is incorrect to use (\ref{link}) inside 1PI diagrams \cite%
{wheelerons}. What we have to do, instead, is apply the strategy explained
in subsection \ref{strat}.

\subsection{Threshold decomposition}

The crucial operation is the threshold decomposition of a diagram, which we
briefly describe here, leaving the details to next subsections.

After integrating on the loop energies by means of the residue theorem, we
obtain the skeletons. Then, we use certain algebraic manipulations to make
all the pseudothresholds $D_{\text{pseudo}}=0$ disappear and only the
physical thresholds $D_{\text{phys}}=0$ survive. At that point we apply the
decomposition%
\begin{equation}
\frac{i}{x+i\epsilon }=\mathcal{P}\frac{i}{x}+\pi \delta (x)  \label{sgnid}
\end{equation}%
and reorganize the results diagrammatically (see below).

Once these operations are completed, we obtain terms proportional $\delta $
functions and terms proportional to products of $\delta $ functions. The
terms proportional to one $\delta $ function are called single thresholds,
while those proportional to a product of more $\delta $ functions are called
multithresholds. When we want to be more precise, we call the latter double
thresholds, triple thresholds, etc., or $\ell $ thresholds, where $\ell $ is
the \textquotedblleft level\textquotedblright\ of the threshold. The support
of the $\delta $ functions is the support of the threshold. Sometimes the
word \textquotedblleft threshold\textquotedblright\ is used indistinctly for
all of them. The \textquotedblleft zero threshold\textquotedblright\
collects the contributions that contain only principal values, which
identify the purely virtual content of the diagram. The zeros of the
principal values are called \textquotedblleft
singularities\textquotedblright .

In the end, the threshold decomposition is the decomposition that properly
organizes and separates the independent thresholds and singularities. We
stress again that they must be physical (no $D_{\text{pseudo}}$ can survive)
and diagrammatic (because generated by diagrams).

The results of the decomposition are often collected in a table (see table %
\ref{ts1} for the triangle) with the following structure. The columns
collect the contributions of the (cut or uncut) skeleton diagrams shown at
the top of them. The rows collect the contributions of the different
(multi)thresholds. The rows are ordered according to the threshold level,
from the zeroth level at the top to the highest level at the bottom. The
smaller tables are shown as equations, while the larger tables are displayed
separately, on top of pages.

The spectral optical identities are the sums of the entries of the rows of a
table. They vanish separately, because each row corresponds to a different
threshold and different thresholds do not interfere with one another. The
spectral optical theorem is obtained by collecting all the entries of the
table in the way explained below. The usual optical theorem is obtained by
integrating the spectral optical theorem (divided by $2\omega $ for every
internal leg) over the phase spaces and the space components of the loop
momenta.

\subsection{Identities for principal values}

We show a simple application of the threshold decomposition just described,
which gives identities that are useful for the calculations of the paper.
Consider the expression%
\begin{equation}
\qquad S(x_{1},\ldots ,x_{n})=\prod\limits_{i=1}^{n}\frac{1}{x_{i}}%
-\sum_{i=1}^{n}\left( \prod\limits_{j=1,j\neq i}^{n}\frac{1}{x_{j}}\right)
\left( \sum_{l=1}^{n}x_{l}\right) ^{-1}.  \label{S1n}
\end{equation}%
If we sum the right-hand side with the least common denominator, we find
zero. This operation is legitimate only if the variables $x_{i}$ and their
sum do not vanish. It is sufficient to assume that they have nonvanishing
imaginary parts $\sigma _{i}$ (typically brought by the infinitesimal widths 
$\epsilon _{a}$) such that $\sum_{i=1}^{n}\sigma _{i}\neq 0$.

It is not legitimate to infer $S(x_{1},\ldots ,x_{n})=0$ if the variables $%
x_{i}$ have real values, in general. In particular, the principal value $%
\mathcal{P}S(x_{1},\ldots ,x_{n})$ may not vanish. To work out its
expression, we start from the identity%
\begin{equation}
0=S(x_{1}+i\epsilon _{1},\ldots ,x_{n}+i\epsilon _{n}),  \label{sn}
\end{equation}%
where the $x_{j}$ are now real and $\epsilon _{j}>0$ for every $j$.
Expanding each term on the right-hand side of (\ref{S1n}) by means of (\ref%
{sgnid}), we can derive its threshold decomposition. The lowest level of the
decomposition is the principal value $\mathcal{P}S(x_{1},\ldots ,x_{n})$ and
the highest level involves only $\delta $ functions. It is easy to show that
(\ref{sn}) gives an identity 
\begin{equation}
\mathcal{P}S(x_{1},\ldots ,x_{n})=c_{n}\pi ^{n}\prod\limits_{i=1}^{n}\delta
(x_{i}),  \label{sn2}
\end{equation}%
where $c_{n}$ is a real constant. We can prove this statement iteratively in 
$n$. For $n=1$ the identities (\ref{sn}) and (\ref{sn2}) are trivial.
Assuming that the statement holds for $n=\hat{n}$, let us consider the case $%
n=\hat{n}+1$. It is easy to show that all the levels of the threshold
decomposition but the lowest and highest ones can be converted into products
of $\delta $ functions by using identities (\ref{sn2}) with $n\leqslant \hat{%
n}$. Thus, the lowest and highest levels must be related by an identity (\ref%
{sn2}) with $n=\hat{n}+1$.

Note that, by reality, $\mathcal{P}S(x_{1},\ldots ,x_{n})=0$ for every odd $%
n $.

Among the identities (\ref{sn2}), we mention the cases $n=2,3,4$, which will
be useful in the next sections. We find%
\begin{eqnarray}
\mathcal{P}\left( \frac{1}{xy}-\frac{1}{x(x+y)}-\frac{1}{y(x+y)}\right) 
&=&-\pi ^{2}\delta (x)\delta (y),  \notag \\
\mathcal{P}\left[ \frac{1}{xyz}-\frac{1}{x+y+z}\left( \frac{1}{xy}+\frac{1}{%
xz}+\frac{1}{yz}\right) \right]  &=&0,  \label{idprin} \\
\mathcal{P}\left[ \frac{1}{xyzw}-\frac{1}{x+y+z+w}\left( \frac{1}{xyz}+\frac{%
1}{xyw}+\frac{1}{xzw}+\frac{1}{yzw}\right) \right]  &=&\pi ^{4}\delta
(x)\delta (y)\delta (z)\delta (w).  \notag
\end{eqnarray}%
A good exercise is to check the steps of the proof outlined above in these
examples. It is also straightforward to verify the identities (\ref{idprin})
numerically, with the help of arbitrary test functions $\varphi
(x_{1},\ldots ,x_{n})$.

The reason why we cannot use common denominators in the expressions between
brackets and conclude that the total is zero is that common denominators
worsen the singularities. Even when the numerators are formally zero, the
total can be nonvanishing.

The identities just found are useful to work out the correct threshold
decomposition. In particular, they allow us to remove all the spurious
thresholds and singularities, which are the non diagrammatic ones and those
due to the denominators $D_{\text{pseudo}}$. The identities must be applied
within the same 1PI diagram. Connected, non 1PI diagrams are treated by
viewing them as products of 1PI subdiagrams (see below).

\subsection{Strategy}

\label{strat}

For future use, it is convenient to define 
\begin{eqnarray}
F^{ab} &=&\frac{1}{e_{a}-e_{b}-\tilde{\omega}_{a}-\tilde{\omega}_{b}},\qquad 
\mathcal{P}^{ab}=\mathcal{P}\frac{1}{e_{a}-e_{b}-\omega _{a}-\omega _{b}}%
,\qquad \mathcal{\hat{P}}^{ab}=\mathcal{P}^{ab}+\mathcal{P}^{ba},  \notag \\
\mathcal{Q}^{ab} &=&\mathcal{P}^{ab}-\mathcal{P}\frac{1}{e_{a}-e_{b}-\omega
_{a}+\omega _{b}},\qquad \qquad \mathcal{\hat{Q}}^{ab}=\left. \mathcal{Q}%
^{ab}\right\vert _{\omega _{a}\rightarrow -\omega _{a}},  \notag \\
\Delta ^{ab} &=&\pi \delta (e_{a}-e_{b}-\omega _{a}-\omega _{b}),
\label{defis}
\end{eqnarray}%
where we recall that $\omega _{a}$ are frequencies and $\tilde{\omega}%
_{a}=\omega _{a}-i\epsilon _{a}$.

It is also useful to introduce some \textquotedblleft
smart\textquotedblright\ manipulations. A \textquotedblleft
smart\textquotedblright\ common denominator identity is an identity of the
form 
\begin{equation}
\sum_{i}c_{i}\prod_{j=1}^{n}\frac{1}{\tilde{D}_{ij}}=\sum_{l}c_{l}^{\prime
}\prod_{j=1}^{n}\frac{1}{\tilde{D}_{\text{phys\hspace{0.01in}}lj}},
\label{smartid}
\end{equation}%
where $c_{i}$ and $c_{l}^{\prime }$ are numerical coefficients and each $%
\tilde{D}_{ij}$ can be a $\tilde{D}_{\text{phys}}$ or a $\tilde{D}_{\text{%
pseudo}}$. Observe that each term of (\ref{smartid}) has the same number $n$
of denominators. Also recall that all the manipulations we make involve
linear combinations of energies and frequencies with coefficients $0$, $+1$
or $-1$. An example of (\ref{smartid}) is the identity%
\begin{equation}
\frac{F^{12}-F^{13}}{e_{2}-e_{3}+\tilde{\omega}_{2}-\tilde{\omega}_{3}}%
=F^{12}F^{13},
\end{equation}%
taken from (\ref{sn}) with $n=2$.

If the left-hand side of (\ref{smartid}) is generated by a diagram, with all
the legs prescribed \`{a} la Feynman, it is always possible to remove the
denominators $\tilde{D}_{\text{pseudo}}$ and obtain a sum of terms like the
right-hand side. This is a key property of the Feynman prescription on the
energies, which guarantees (perturbative) stability.

\bigskip

The strategy of the calculations is as follows. We assume, for simplicity,
that $G$ is one-particle irreducible, the extension to connected, non 1PI
diagrams being straightforward.

1) Start from the diagram $G$, prescribed \`{a} la Feynman.

2) Integrate on the loop energies by means of the residue theorem.

3) Use smart common-denominator identities (\ref{smartid}) to remove all the
denominators $\tilde{D}_{\text{pseudo}}$ and leave only denominators $\tilde{%
D}_{\text{phys}}$.

4) Perform the threshold decomposition as follows:

4.a) separate each fraction $1/\tilde{D}_{\text{phys}}$ into its purely
virtual part and its on-shell part by means of (\ref{sgnid});

4.b) use identities for principal values such as (\ref{idprin}) to arrange
the decomposition in a proper diagrammatic form (see below for the meaning
of this).

5) Perform the fakeon projection, by eliminating all the thresholds that
involve fakeon frequencies (which will be called \textquotedblleft fakeon
thresholds\textquotedblright ).

\bigskip

Steps 1) to 4) lead to the spectral optical identities of the diagrams where
each leg is prescribed \`{a} la Feynman. If some legs are prescribed as
fakeons, the identities reduce to a proper subset, as per step 5).

The purely virtual content of a diagram is its level 0. It is associated
with the diagram where all the internal legs are fakeons. Away from the
thresholds, it coincides with the Euclidean version of the skeleton. It
admits no nontrivial cuts.

Due to the diagrammatic structure mentioned in point 4.b), all the threshold
contributions (level $\geqslant 1$) are manifestly Lorentz invariant, once
we integrate on the space components of the loop momenta and the phase
spaces. By subtraction, the purely virtual part is also Lorentz invariant,
although not manifestly.

The threshold decomposition\ also applies to cut diagrams. There, step 2) is
simplified by the presence of cut propagators, which fix some loop energies
without the need to use the residue theorem.

It remains to explain what we mean by step 4.b) and why it is always
possible to achieve the proper diagrammatic form it refers to.

\subsection{Proper diagrammatic form and proper decomposition}

Before defining the proper diagrammatic form of the threshold decomposition,
we illustrate the problem with two examples at one loop, level $\ell =1$:
the product $\Delta ^{12}\mathcal{P}^{32}\mathcal{P}^{34}$, which does not
have a proper diagrammatic form, and the product $\Delta ^{12}\mathcal{P}%
^{32}\mathcal{P}^{14}$, which does. If we reinstate the loop energy $k^{0}$
leading to $\Delta ^{12}$, we find%
\begin{eqnarray}
\Delta ^{12}\mathcal{P}^{32}\mathcal{P}^{34} &=&2\pi ^{2}\mathcal{P}\int 
\frac{\mathrm{d}k^{0}}{2\pi }\delta (k^{0}-e_{1}+\omega _{1})\delta
(k^{0}-e_{2}-\omega _{2})\frac{1}{e_{3}-k^{0}-\omega _{3}}\frac{1}{%
e_{3}-e_{4}-\omega _{3}-\omega _{4}},  \notag \\
\Delta ^{12}\mathcal{P}^{32}\mathcal{P}^{14} &=&2\pi ^{2}\mathcal{P}\int 
\frac{\mathrm{d}k^{0}}{2\pi }\delta (k^{0}-e_{1}+\omega _{1})\delta
(k^{0}-e_{2}-\omega _{2})\frac{1}{e_{3}-k^{0}-\omega _{3}}\frac{1}{%
k^{0}-e_{4}-\omega _{4}}.  \label{reinste}
\end{eqnarray}%
It is evident that the second line has a proper diagrammatic form, with the
loop energy circulating in every factor. In the first line, the $\delta $
function of $\Delta ^{12}$ cannot be used to modify $\mathcal{P}^{34}$. The
reason is that the skeleton $\Delta ^{12}\mathcal{P}^{32}\mathcal{P}^{34}$
has a non diagrammatic singularity, obtained by combining the singularities
of $\Delta ^{12}$ and $\mathcal{P}^{34}$: 
\begin{equation}
e_{1}-e_{2}-\omega _{1}-\omega _{2}=e_{3}-e_{4}-\omega _{3}-\omega _{4}=0.
\label{nondiaga}
\end{equation}%
Such a singularity is not present in the skeletons we want to decompose,
since they are originated by diagrams. This means that it must cancel with
an opposite contribution coming from some other source. In section \ref{boxG}
we show that indeed it does.

The manipulations advocated in step 4.b) remove contributions like $\Delta
^{12}\mathcal{P}^{32}\mathcal{P}^{34}$ in favor of contributions like $%
\Delta ^{12}\mathcal{P}^{32}\mathcal{P}^{14}$. The importance of operation
4.b) will be appreciated starting from the box (section \ref{boxG}), since
the triangle is too simple\ in this respect.

\bigskip

The proper form of the threshold decomposition (proper decomposition, from
now on) of an $L$ loop connected diagram (prescribed \`{a} la Feynman
everywhere) is the product of the proper decompositions of its 1PI
ingredients. Vertices are considered 1PI\ irreducible subdiagrams. A
propagator with no vertices attached to it is viewed as an 1PI\ subdiagram
as well. The proper decomposition of a tree diagram follows from the
diagrammatic rules of subsection \ref{diagra} (every propagator being a
link, in the sense explained there).

The proper decomposition of an $L$ loop 1PI diagram is defined by induction.
Consider a 1PI skeleton diagram $\mathcal{G}_{L}^{s}$ with $L$ loops and $I$
internal legs. Let%
\begin{equation*}
\mathcal{G}_{L}^{s}=\sum_{\ell =0}^{I-L}\mathcal{G}_{L,\mathcal{\ell }}^{s}
\end{equation*}%
denote its threshold decomposition, organized into levels $\ell $. The $\ell
=0$ contribution $\mathcal{G}_{L,0}^{s}$ is the only one that truly has $L$
loops, because every $\mathcal{G}_{L,\mathcal{\ell }}^{s}$ with $\ell
\geqslant 1$ carries at least one $\delta $ function, which effectively
breaks the skeleton by interrupting an internal leg and setting a condition
on the external momenta. In the case of cut diagrams, the $\ell =0$
contribution $\mathcal{G}_{L,0}^{s}$ is absent.

The proper decomposition of an $L$-loop cut diagram is straightforward,
because a cut diagram is broken by the cuts into subdiagrams with smaller
numbers of loops (whose proper decompositions are already known by the
inductive assumption).

The proper decomposition of an uncut diagram is determined as follows. The
decomposition of $\mathcal{G}_{L,\mathcal{\ell }}^{s}$ is known from the
optical theorem for every odd $\mathcal{\ell }$, because it must match the
decompositions of the cut diagrams. The decomposition of each $\mathcal{G}%
_{L,\mathcal{\ell }}^{s}$ with even $\ell \geqslant 2$ is worked out from
the one of $\mathcal{G}_{L,\ell -1}^{s}$ by circumventing the singularities
associated with the principal values as demanded by the Feynman prescription
(and paying attention to the combinatorics, to avoid overcounting). There is
no ambiguity in this, since identities like (\ref{idprin}) relate levels
separated by an even number. Finally, the decomposition of $\mathcal{G}%
_{L,0}^{s}$ is determined by subtraction.

\bigskip

The proper diagrammatic form of the zeroth level $\mathcal{G}_{L,0}^{s}$ is
not known a priori, but determined by the procedure itself. The reason is
that the zeroth level is a sum of residues where every loop energy $k^{0}$
can have different values: several spurious thresholds and singularities
(like the non diagrammatic ones and those due to the denominators $D_{\text{%
pseudo}}$) mutually cancel, although it is not manifest that they do.

The proper diagrammatic form of the threshold decomposition always exists
and is unique. The skeleton singularities have a diagrammatic origin, so the
non diagrammatic singularities can only appear in the intermediate steps,
especially due to the manipulations of step 3), but must cancel out in the
total. Moreover, the thresholds and singularities must appear with the right
coefficients to match the diagram $G$ they are generated by. When that
happens, the thresholds are properly extracted from the zeroth level to the
levels $\ell \geqslant 1$ and the zeroth level defines the purely virtual
content of the diagram $G$.

\section{Bubbles}

\label{bubbles}\setcounter{equation}{0}

In this section we study the (multi)bubble diagrams. Although they do not
present particular difficulties, they are a useful guide through the more
complex configurations we study in the next sections.

The simple bubble diagram $\rangle \hspace{-0.18em}{\bigcirc \hspace{-0.18em}%
\langle }$ with physical particles propagating in the internal legs is
denoted by (AB), or (A1B2). It gives the skeleton $B_{\text{AB}%
}^{s}=G_{2}^{s}$ of formula (\ref{GsN}). Integrating on the loop energy $%
k^{0}$ by means of the residue theorem, we obtain%
\begin{equation}
B_{\text{AB}}^{s}=-i\left( F^{12}-F_{+-}^{12}\right) -i\left(
F^{21}-F_{+-}^{21}\right) ,  \label{FF}
\end{equation}%
where $F_{+-}^{ab}=\left. F^{ab}\right\vert _{\omega _{a}\rightarrow -\omega
_{a}}$. While $F^{12}$ and $F^{21}$ have the form $1/\tilde{D}_{\text{phys}}$%
, $F_{+-}^{12}$ and $F_{+-}^{21}$ have the form $1/\tilde{D}_{\text{pseudo}}$%
. As expected, the last two mutually cancel (with no need to use the
identities (\ref{smartid}), in this case). In the end, we get%
\begin{equation}
B_{\text{AB}}^{s}=-i(F^{12}+F^{21})=-\frac{2i(\omega _{1}+\omega _{2})}{%
(e_{1}-e_{2})^{2}-(\omega _{1}+\omega _{2})^{2}+i\epsilon }.  \label{tb0}
\end{equation}

Using the decomposition (\ref{sgnid}) for $B_{\text{AB}}^{s}$, which here
reads%
\begin{equation}
F^{ab}=\mathcal{P}^{ab}-i\Delta ^{ab},  \label{sgnid2}
\end{equation}%
we find%
\begin{equation}
B_{\text{AB}}^{s}=-i\mathcal{\hat{P}}^{12}-\Delta ^{12}-\Delta ^{21}.
\end{equation}%
The conjugate diagram is $B_{\text{\.{A}\.{B}}}^{s}=\bar{B}_{\text{AB}}^{s}$%
. From the rules (\ref{propag}), the cut diagrams are $B_{\text{\.{A}B}%
}^{s}=2\Delta ^{21}$ and $B_{\text{A\.{B}}}^{s}=2\Delta ^{12}$ (orienting $%
p_{1}$ from A to B). Collecting everything together, we obtain the table 
\begin{equation}
\begin{tabular}{|c|c|c|c|c|}
\hline
Th$\setminus $G & $B_{\text{AB}}^{s}$ & $B_{\text{\.{A}\.{B}}}^{s}$ & $B_{%
\text{\.{A}B}}^{s}$ & $B_{\text{A\.{B}}}^{s}$ \\ \hline
--- & $-i\mathcal{\hat{P}}^{12}$ & $i\mathcal{\hat{P}}^{12}$ & $0$ & $0$ \\ 
\hline
$\Delta ^{12}$ & $-1$ & $-1$ & $0$ & $2$ \\ \hline
$\Delta ^{21}$ & $-1$ & $-1$ & $2$ & $0$ \\ \hline
\end{tabular}
\label{tb1}
\end{equation}%
where \textquotedblleft Th\textquotedblright\ stands for threshold and
\textquotedblleft G\textquotedblright\ stands for diagram. The thresholds
are listed vertically, while the diagrams are listed horizontally.

Let $C_{ij}$ denote the entries of the table. A (cut or uncut) diagram $G_{j}
$ is the $j$th column of the table ($j>1$), by which we mean the sum%
\begin{equation}
G_{j}\equiv \sum_{i>1}C_{i1}C_{ij},  \label{columns}
\end{equation}%
where $C_{21}=1$. The spectral optical identities are \textquotedblleft the
rows of the table\textquotedblright , by which we mean the identities%
\begin{equation}
R_{i}\equiv C_{i1}\sum_{j>1}C_{ij}=0,  \label{rows}
\end{equation}%
for $i>1$. The spectral optical theorem 
\begin{equation}
B_{\text{AB}}^{s}+B_{\text{\.{A}\.{B}}}^{s}+B_{\text{\.{A}B}}^{s}+B_{\text{A%
\.{B}}}^{s}=0  \label{SOTb}
\end{equation}%
is the \textquotedblleft sum of the entries\textquotedblright\ of (\ref{tb1}%
), by which we mean the identity%
\begin{equation}
\sum_{j>1}G_{j}=\sum_{i>1}\sum_{j>1}C_{i1}C_{ij}=0.  \label{spoct}
\end{equation}%
Finally, the optical theorem is the integral of this identity, multiplied by 
$4\omega _{1}\omega _{2}$, over the space components of the loop momentum.

With a fakeon in either internal leg, or in both, we denote the diagram by
(AfB), (AfBf), etc., depending on the case. We need to apply step 5) of
subsection \ref{strat}, which amounts to drop the thresholds identified by $%
\Delta ^{12}$ and $\Delta ^{21}$, since each of them contains at least one
fakeon frequency. The table loses the last two rows, so we obtain%
\begin{equation}
B_{\text{AfB}}^{s}=B_{\text{AfBf}}^{s}=-i\mathcal{\hat{P}}^{12},\qquad B_{%
\text{\.{A}f\.{B}}}^{s}=B_{\text{\.{A}f\.{B}f}}^{s}=i\mathcal{\hat{P}}^{12}.
\label{tb2}
\end{equation}%
No nontrivial cut diagram survives, in agreement with the rules of
subsection \ref{diagra}.

The double bubble $\rangle \hspace{-0.18em}{\bigcirc \hspace{-0.18em}{|%
\hspace{-0.18em}{\bigcirc \hspace{-0.18em}\langle }}}$ is (AB)(BC), or
(A1B2)(B3C4). Without the risk of confusion, we just call it ABC\ in table %
\ref{tb3}, where we show its threshold decomposition. 
\begin{table}[t]
\begin{center}
\begin{tabular}{|c|c|c|c|c|c|c|c|c|}
\hline
Th$\setminus $G & $B_{\text{ABC}}^{s}$ & $B_{\text{\.{A}\.{B}\.{C}}}^{s}$ & $%
B_{\text{\.{A}BC}}^{s}$ & $B_{\text{A\.{B}C}}^{s}$ & $B_{\text{AB\.{C}}}^{s}$
& $B_{\text{A\.{B}\.{C}}}^{s}$ & $B_{\text{\.{A}B\.{C}}}^{s}$ & $B_{\text{\.{%
A}\.{B}C}}^{s}$ \\ \hline
--- & $-i\mathcal{\hat{P}}^{12}\mathcal{\hat{P}}^{34}$ & $i\mathcal{\hat{P}}%
^{12}\mathcal{\hat{P}}^{34}$ & $0$ & $0$ & $0$ & $0$ & $0$ & $0$ \\ \hline
$\Delta ^{12}$ & $-\mathcal{\hat{P}}^{34}$ & $-\mathcal{\hat{P}}^{34}$ & $0$
& $0$ & $0$ & $2\mathcal{\hat{P}}^{34}$ & $0$ & $0$ \\ \hline
$\Delta ^{21}$ & $-\mathcal{\hat{P}}^{34}$ & $-\mathcal{\hat{P}}^{34}$ & $2%
\mathcal{\hat{P}}^{34}$ & $0$ & $0$ & $0$ & $0$ & $0$ \\ \hline
$\Delta ^{34}$ & $-\mathcal{\hat{P}}^{12}$ & $-\mathcal{\hat{P}}^{12}$ & $0$
& $0$ & $2\mathcal{\hat{P}}^{12}$ & $0$ & $0$ & $0$ \\ \hline
$\Delta ^{43}$ & $-\mathcal{\hat{P}}^{12}$ & $-\mathcal{\hat{P}}^{12}$ & $0$
& $0$ & $0$ & $0$ & $0$ & $2\mathcal{\hat{P}}^{12}$ \\ \hline
$\Delta ^{12}\Delta ^{34}$ & $i$ & $-i$ & $0$ & $0$ & $-2i$ & $2i$ & $0$ & $%
0 $ \\ \hline
$\Delta ^{12}\Delta ^{43}$ & $i$ & $-i$ & $0$ & $-4i$ & $0$ & $2i$ & $0$ & $%
2i$ \\ \hline
$\Delta ^{21}\Delta ^{34}$ & $i$ & $-i$ & $-2i$ & $0$ & $-2i$ & $0$ & $4i$ & 
$0$ \\ \hline
$\Delta ^{21}\Delta ^{43}$ & $i$ & $-i$ & $-2i$ & $0$ & $0$ & $0$ & $0$ & $%
2i $ \\ \hline
\end{tabular}%
\end{center}
\caption{Threshold decomposition of the double bubble}
\label{tb3}
\end{table}

With a fakeon in leg 1, we drop $\Delta ^{12}$ and $\Delta ^{21}$ everywhere
and obtain%
\begin{equation}
\begin{tabular}{|c|c|c|c|c|}
\hline
Th$\setminus $G & $B_{\text{AfBC}}^{s}$ & $B_{\text{\.{A}f\.{B}\.{C}}}^{s}$
& $B_{\text{AfB\.{C}}}^{s}$ & $B_{\text{\.{A}f\.{B}C}}^{s}$ \\ \hline
--- & $-i\mathcal{\hat{P}}^{12}\mathcal{\hat{P}}^{34}$ & $i\mathcal{\hat{P}}%
^{12}\mathcal{\hat{P}}^{34}$ & $0$ & $0$ \\ \hline
$\Delta ^{34}$ & $-\mathcal{\hat{P}}^{12}$ & $-\mathcal{\hat{P}}^{12}$ & $2%
\mathcal{\hat{P}}^{12}$ & $0$ \\ \hline
$\Delta ^{43}$ & $-\mathcal{\hat{P}}^{12}$ & $-\mathcal{\hat{P}}^{12}$ & $0$
& $2\mathcal{\hat{P}}^{12}$ \\ \hline
\end{tabular}
\label{tb4}
\end{equation}

With fakeons in legs 1 and 2, we get the same. With a fakeon in leg 1 and
one in leg 3, we obtain%
\begin{equation}
B_{\text{AfBfC}}^{s}=-i\mathcal{\hat{P}}^{12}\mathcal{\hat{P}}^{34},\qquad
B_{\text{\.{A}f\.{B}f\.{C}}}^{s}=i\mathcal{\hat{P}}^{12}\mathcal{\hat{P}}%
^{34}.  \label{tb5}
\end{equation}

We can repeat the derivation for the skeletons of multibubble diagrams $%
\rangle \hspace{-0.18em}{\bigcirc \hspace{-0.18em}{|\hspace{-0.18em}{%
\bigcirc \hspace{-0.18em}{|\hspace{-0.18em}{\bigcirc \hspace{-0.18em}{|%
\hspace{-0.18em}{\bigcirc \hspace{-0.18em}{|\hspace{-0.18em}{\bigcirc 
\hspace{-0.18em}\langle }}}}}}}}}$. The result is the product of the
skeletons of the single bubbles, for all choices of physical particles and
fakeons.

In all cases, the sum (\ref{columns}) of each column equals the (cut or
uncut) skeleton diagram shown at the top of it. Moreover, the spectral
optical identities are the sums (\ref{rows}) of the rows. The spectral
optical theorem is obtained by summing all the entries of the table as in (%
\ref{spoct}). For example, for the double bubble diagram $\rangle \hspace{%
-0.18em}{\bigcirc \hspace{-0.18em}{|\hspace{-0.18em}{\bigcirc \hspace{-0.18em%
}\langle }}}$ we have%
\begin{eqnarray}
&&B_{\text{ABC}}^{s}+B_{\text{\.{A}\.{B}\.{C}}}^{s}+B_{\text{\.{A}BC}%
}^{s}+B_{\text{A\.{B}C}}^{s}+B_{\text{AB\.{C}}}^{s}+B_{\text{A\.{B}\.{C}}%
}^{s}+B_{\text{\.{A}B\.{C}}}^{s}+B_{\text{\.{A}\.{B}C}}^{s}=0,\quad \text{%
with no fakeons,}  \notag \\
&&B_{\text{AfBC}}^{s}+B_{\text{\.{A}f\.{B}\.{C}}}^{s}+B_{\text{AfB\.{C}}%
}^{s}+B_{\text{\.{A}f\.{B}C}}^{s}=0,\quad \text{with a fakeon in AB,}  \notag
\\
&&B_{\text{AfBfC}}^{s}+B_{\text{\.{A}f\.{B}f\.{C}}}^{s}=0,\quad \text{with a
fakeon in AB\ and one in BC.}
\end{eqnarray}

The usual optical theorem is obtained by integrating the spectral optical
theorem (divided by $2\omega $ for every internal leg) over the space
components of the loop momentum or the phase spaces (in cut diagrams). For
example, with a fakeon in AB, we obtain%
\begin{equation}
B_{\text{AfBC}}+B_{\text{\.{A}f\.{B}\.{C}}}+B_{\text{AfB\.{C}}}+B_{\text{\.{A%
}f\.{B}C}}=0.
\end{equation}

The final outcome is that the single bubble diagram with one or two internal
fakeons is equal to $i$ times the imaginary part of the usual bubble diagram:%
\begin{equation}
\resizebox{5cm}{!}{$ \overset{\text{f}}{\rangle \hspace{-0.18em}{\bigcirc
\hspace{-0.18em}\langle }}=\underset{\text{f}}{\overset{\text{f}}{\rangle
\hspace{-0.18em}{\bigcirc \hspace{-0.18em}\langle }}}=i\,\text{Im}\left[
\rangle \hspace{-0.18em}{\bigcirc \hspace{-0.18em}\langle }\right] $}\text{.}
\end{equation}%
The $n$th multibubble diagram is the product of the single bubble diagrams
it is made of, times $i^{n-1}$. It is easy to check that the diagrammatics
of cut and uncut diagrams, with or without fakeons, agrees with the one
stated in subsection \ref{diagra}.

\section{Triangle}

\label{triangleT}\setcounter{equation}{0}

In this section we study the triangle diagram, denoted by (ABC), or
(A1B2C3). Its skeleton is $T_{\text{ABC}}^{s}=G_{3}^{s}$, from formula (\ref%
{GsN}). For convenience, we integrate on the loop energy $k^{0}$ by
averaging on the two ways to close the integration path at infinity and
using the residue theorem. Without using common denominators or other
manipulations, the result can be expanded as 
\begin{equation}
T_{\text{ABC}}^{s}=-iF_{\text{ABC}}-iF_{1\text{ABC}}-iF_{2\text{ABC}},
\label{ts2}
\end{equation}%
where%
\begin{eqnarray}
F_{\text{ABC}} &=&F^{12}F^{13}+\text{cycl}+(e\rightarrow -e),  \notag \\
F_{1\text{ABC}} &=&\frac{1}{2}S^{12|13}+\text{cycl}+(e\rightarrow -e),\qquad
F_{2\text{ABC}}=-\frac{1}{2}S_{++}^{12|31}+(e\rightarrow -e), \\
S^{ab|cd} &=&S(e_{a}-e_{b}-\tilde{\omega}_{a}-\tilde{\omega}%
_{b},-e_{c}+e_{d}+\tilde{\omega}_{c}+\tilde{\omega}_{d}),\qquad \hat{S}%
^{ab|cd}=\left. S^{ab|dc}\right\vert _{\tilde{\omega}_{b}\rightarrow -\tilde{%
\omega}_{b},\ \tilde{\omega}_{c}\rightarrow -\tilde{\omega}_{c}}.  \notag
\end{eqnarray}%
Here, \textquotedblleft cycl\textquotedblright\ refers to the cyclic
permutations of 1, 2 and 3 and $S(x,y)$ is defined in (\ref{S1n}).

Because of identities like (\ref{sn}), both $S^{ab|cd}$ and $\hat{S}^{ab|cd}$
vanish. This is where the operation (\ref{smartid}) eliminates all the
pseudothresholds and leaves the physical thresholds only. We obtain%
\begin{equation}
T_{\text{ABC}}^{s}=-iF_{\text{ABC}}.  \label{ts3}
\end{equation}%
With the help of (\ref{sgnid}), or (\ref{sgnid2}), the threshold
decomposition gives%
\begin{equation}
T_{\text{ABC}}^{s}=-i\mathcal{P}_{\text{ABC}}-\sum_{\text{perms}}\Delta ^{ab}%
\mathcal{Q}^{ac}+\frac{i}{2}\sum_{\text{perms}}\Delta ^{ab}(\Delta
^{ac}+\Delta ^{cb}),  \label{Tdecomp}
\end{equation}%
where $\mathcal{P}_{\text{ABC}}=\left. F_{\text{ABC}}\right\vert
_{F\rightarrow \mathcal{P}}$ and the sums are on $\{a,b,c\}$ equal to the
permutations of 1, 2 and 3. The conjugate diagram is $T_{\text{\.{A}\.{B}%
\.{C}}}^{s}=\bar{T}_{\text{ABC}}^{s}$. The cut diagrams follow from the
diagrammatic rules of subsection \ref{diagra}. Orienting $p_{1}$ from B to
A, we have 
\begin{equation}
T_{\text{A\.{B}C}}^{s}=2\Delta ^{21}(\mathcal{Q}^{23}-i\Delta ^{31}-i\Delta
^{23}),\qquad T_{\text{\.{A}B\.{C}}}^{s}=2\Delta ^{12}(\mathcal{Q}%
^{13}+i\Delta ^{13}+i\Delta ^{32}),
\end{equation}%
the other ones being derived by cyclically permuting 1, 2 and 3. In the end,
we obtain table \ref{ts1}. The diagrams are given by the columns, as per (%
\ref{columns}). 
\begin{table}[t]
\begin{center}
\begin{tabular}{|c|c|c|c|c|c|c|c|c|}
\hline
Th$\setminus $G & $T_{\text{ABC}}^{s}$ & $T_{\text{\.{A}\.{B}\.{C}}}^{s}$ & $%
T_{\text{\.{A}BC}}^{s}$ & $T_{\text{A\.{B}C}}^{s}$ & $T_{\text{AB\.{C}}}^{s}$
& $T_{\text{A\.{B}\.{C}}}^{s}$ & $T_{\text{\.{A}B\.{C}}}^{s}$ & $T_{\text{\.{%
A}\.{B}C}}^{s}$ \\ \hline
--- & $-i\mathcal{P}_{\text{ABC}}$ & $i\mathcal{P}_{\text{ABC}}$ & $0$ & $0$
& $0$ & $0$ & $0$ & $0$ \\ \hline
$\Delta ^{23}$ & $-\mathcal{Q}^{21}$ & $-\mathcal{Q}^{21}$ & $0$ & $0$ & $0$
& $0$ & $0$ & $2\mathcal{Q}^{21}$ \\ \hline
$\Delta ^{12}$ & $-\mathcal{Q}^{13}$ & $-\mathcal{Q}^{13}$ & $0$ & $0$ & $0$
& $0$ & $2\mathcal{Q}^{13}$ & $0$ \\ \hline
$\Delta ^{31}$ & $-\mathcal{Q}^{32}$ & $-\mathcal{Q}^{32}$ & $0$ & $0$ & $0$
& $2\mathcal{Q}^{32}$ & $0$ & $0$ \\ \hline
$\Delta ^{32}$ & $-\mathcal{Q}^{31}$ & $-\mathcal{Q}^{31}$ & $0$ & $0$ & $2%
\mathcal{Q}^{31}$ & $0$ & $0$ & $0$ \\ \hline
$\Delta ^{21}$ & $-\mathcal{Q}^{23}$ & $-\mathcal{Q}^{23}$ & $0$ & $2%
\mathcal{Q}^{23}$ & $0$ & $0$ & $0$ & $0$ \\ \hline
$\Delta ^{13}$ & $-\mathcal{Q}^{12}$ & $-\mathcal{Q}^{12}$ & $2\mathcal{Q}%
^{12}$ & $0$ & $0$ & $0$ & $0$ & $0$ \\ \hline
$\Delta ^{12}\Delta ^{13}$ & $i$ & $-i$ & $-2i$ & $0$ & $0$ & $0$ & $2i$ & $%
0 $ \\ \hline
$\Delta ^{23}\Delta ^{21}$ & $i$ & $-i$ & $0$ & $-2i$ & $0$ & $0$ & $0$ & $%
2i $ \\ \hline
$\Delta ^{31}\Delta ^{32}$ & $i$ & $-i$ & $0$ & $0$ & $-2i$ & $2i$ & $0$ & $%
0 $ \\ \hline
$\Delta ^{31}\Delta ^{21}$ & $i$ & $-i$ & $0$ & $-2i$ & $0$ & $2i$ & $0$ & $%
0 $ \\ \hline
$\Delta ^{12}\Delta ^{32}$ & $i$ & $-i$ & $0$ & $0$ & $-2i$ & $0$ & $2i$ & $%
0 $ \\ \hline
$\Delta ^{23}\Delta ^{13}$ & $i$ & $-i$ & $-2i$ & $0$ & $0$ & $0$ & $0$ & $%
2i $ \\ \hline
\end{tabular}%
\end{center}
\caption{Threshold decomposition of the triangle}
\label{ts1}
\end{table}

The entries of the table are organized in a diagrammatic form, as per point
4.b) of subsection \ref{strat}. Indeed, 
\begin{equation}
\Delta ^{ab}\mathcal{Q}^{ac}=\Delta ^{ab}\mathcal{Q}^{cb}=16\pi ^{2}\omega
_{a}\omega _{b}\omega _{c}\mathcal{P}\int \frac{\mathrm{d}k^{0}}{2\pi }\frac{%
\delta ^{-}((k-p_{a})^{2}-m_{a}^{2})\delta ^{+}((k-p_{b})^{2}-m_{b}^{2})}{%
(k-p_{c})^{2}-m_{c}^{2}},  \label{DeltaP}
\end{equation}%
where $\delta ^{\pm }(p^{2}-m^{2})=\theta (\pm p^{0})\delta (p^{2}-m^{2})$.
As soon as we divide by $8\omega _{a}\omega _{b}\omega _{c}$ and integrate
on the space components of the loop momentum, the result is Lorentz
invariant.

The spectral optical identities are the rows (\ref{rows}) of table \ref{ts1}%
, which vanish separately. The spectral optical theorem is obtained by
summing all the entries of the table as in (\ref{spoct}):%
\begin{equation}
T_{\text{ABC}}^{s}+T_{\text{\.{A}\.{B}\.{C}}}^{s}+T_{\text{\.{A}BC}}^{s}+T_{%
\text{A\.{B}C}}^{s}+T_{\text{AB\.{C}}}^{s}+T_{\text{A\.{B}\.{C}}}^{s}+T_{%
\text{\.{A}B\.{C}}}^{s}+T_{\text{\.{A}\.{B}C}}^{s}=0.  \label{SOTt}
\end{equation}%
The usual optical theorem 
\begin{equation}
T_{\text{ABC}}+T_{\text{\.{A}\.{B}\.{C}}}+T_{\text{\.{A}BC}}+T_{\text{A\.{B}C%
}}+T_{\text{AB\.{C}}}+T_{\text{A\.{B}\.{C}}}+T_{\text{\.{A}B\.{C}}}+T_{\text{%
\.{A}\.{B}C}}=0  \label{SOT}
\end{equation}%
is obtained dividing (\ref{SOTt}) by $8\omega _{1}\omega _{2}\omega _{3}$
and integrating on the space components $\mathbf{k}$ of the loop momentum.
It is also evident that (\ref{SOTt}) and (\ref{SOT}) agree with the
diagrammatic rules of subsection \ref{diagra}.

\subsection{Fakeons}

Now we study the fakeon prescription/projection. Assume that leg 1, which is
the segment AB, is a fakeon and the other two internal legs are physical
particles. Then $\omega _{1}$ is a fakeon frequency. According to step 5) of
subsection \ref{strat}, we must suppress all the thresholds involving $%
\omega _{1}$, i.e., the single thresholds proportional to $\Delta ^{12}$, $%
\Delta ^{21}$, $\Delta ^{13}$ and $\Delta ^{31}$, and all the double
thresholds. So doing, we obtain 
\begin{eqnarray}
T_{\text{AfBC}}^{s} &=&-i\mathcal{P}_{\text{ABC}}-\Delta ^{23}\mathcal{Q}%
^{21}-\Delta ^{32}\mathcal{Q}^{31},\qquad T_{\text{AfB\.{C}}}^{s}=2\Delta
^{32}\mathcal{Q}^{31},  \notag \\
T_{\text{\.{A}f\.{B}\.{C}}}^{s} &=&i\mathcal{P}_{\text{ABC}}-\Delta ^{23}%
\mathcal{Q}^{21}-\Delta ^{32}\mathcal{Q}^{31},\qquad T_{\text{\.{A}f\.{B}C}%
}^{s}=2\Delta ^{23}\mathcal{Q}^{21},  \label{Tcut}
\end{eqnarray}%
and the table 
\begin{equation}
\begin{tabular}{|c|c|c|c|c|}
\hline
Th$\setminus $G & $T_{\text{AfBC}}^{s}$ & $T_{\text{\.{A}f\.{B}\.{C}}}^{s}$
& $T_{\text{AfB\.{C}}}^{s}$ & $T_{\text{\.{A}f\.{B}C}}^{s}$ \\ \hline
--- & $-i\mathcal{P}_{\text{ABC}}$ & $i\mathcal{P}_{\text{ABC}}$ & $0$ & $0$
\\ \hline
$\Delta ^{23}$ & $-\mathcal{Q}^{21}$ & $-\mathcal{Q}^{21}$ & $0$ & $2%
\mathcal{Q}^{21}$ \\ \hline
$\Delta ^{32}$ & $-\mathcal{Q}^{31}$ & $-\mathcal{Q}^{31}$ & $2\mathcal{Q}%
^{31}$ & $0$ \\ \hline
\end{tabular}%
\end{equation}%
As usual, the diagrams are the columns (\ref{columns}) of the table and the
spectral optical identities are the rows (\ref{rows}). The spectral optical
theorem and the ordinary optical theorem are the sums%
\begin{eqnarray}
T_{\text{AfBC}}^{s}+T_{\text{\.{A}f\.{B}\.{C}}}^{s}+T_{\text{AfB\.{C}}%
}^{s}+T_{\text{\.{A}f\.{B}C}}^{s} &=&0,  \notag \\
T_{\text{AfBC}}+T_{\text{\.{A}f\.{B}\.{C}}}+T_{\text{AfB\.{C}}}+T_{\text{\.{A%
}f\.{B}C}} &=&0.
\end{eqnarray}%
Again, these identities agree with the diagrammatic rules of section \ref%
{diagra}. In particular, since the shadowed and unshadowed portions of the
cut triangle are tree diagrams, the fakeon propagator in AB is a link in $T_{%
\text{AfB\.{C}}}^{s}$ and $T_{\text{\.{A}f\.{B}C}}^{s}$, so it is given by (%
\ref{link}).

With two or three fakeons, we drop all the thresholds and obtain%
\begin{equation}
T_{\text{AfBfC}}^{s}=T_{\text{AfBfCf}}^{s}=-i\mathcal{P}_{\text{ABC}}=i%
\hspace{0.01in}\text{Im[}T_{\text{ABC}}^{s}\text{]}-\frac{i}{2}\sum_{\text{%
perms}}\Delta ^{ab}(\Delta ^{ac}+\Delta ^{cb}).  \label{TPV}
\end{equation}%
In this case, the skeleton diagram is purely imaginary, so no cut diagrams
survive. The result (\ref{TPV}) encodes the purely virtual content of the
triangle.

Formulas (\ref{Tcut}) and (\ref{TPV})\ can be used to relate the triangles
with fakeons to the standard triangle, for possible implementations in
softwares like FeynCalc, FormCalc, LoopTools and Package-X \cite{calc}.

\section{Box}

\label{boxG}\setcounter{equation}{0}

In this section we study the box diagram $G_{4}$, denoted by (ABCD), or
(A1B2C3D4). We consider its skeleton $G_{4}^{s}$ and integrate on the loop
energy $k^{0}$. Then, we use smart identities (\ref{smartid}) to make the
pseudothresholds disappear, which we know to be possible because they are
not physical. The result of these operations is%
\begin{equation}
G_{\text{ABCD}}^{s4}=-\frac{i}{6}\sum_{\text{perms}}F^{ab}F^{ac}F^{ad}-\frac{%
i}{4}\sum_{\text{perms}}F^{ab}F^{ac}F^{db}+(e\rightarrow -e),  \label{G4s}
\end{equation}%
where the sums are evaluated for $\{a,b,c,d\}$ ranging over the set of
permutations of 1, 2, 3 and 4.

To derive the threshold decomposition, we first apply (\ref{sgnid}), or (\ref%
{sgnid2}), and expand. The terms proportional one $\delta $ function, which
are%
\begin{equation}
-\frac{1}{2}\sum_{\text{perms}}\Delta ^{ab}\mathcal{P}^{ac}\mathcal{P}^{ad}-%
\frac{1}{4}\sum_{\text{perms}}(\Delta ^{ab}\mathcal{P}^{ac}\mathcal{P}^{db}+%
\mathcal{P}^{ab}\Delta ^{ac}\mathcal{P}^{db}+\mathcal{P}^{ab}\mathcal{P}%
^{ac}\Delta ^{db})
\end{equation}%
plus $(e\rightarrow -e)$, mix single thresholds and triple thresholds, due
to identities like (\ref{idprin}). We can separate the two types of
contributions by fulfilling the proper diagrammatic requirement 4.b) of
subsection \ref{strat}. Consider, for example, the terms proportional to $%
\Delta ^{12}$ (all the others being treated the same way), which read%
\begin{eqnarray}
w_{12} &\equiv &-\Delta ^{12}\left[ \mathcal{P}^{13}\mathcal{P}^{14}+%
\mathcal{P}^{32}\mathcal{P}^{42}+\frac{1}{2}(\mathcal{P}^{14}\mathcal{P}%
^{32}+\mathcal{P}^{14}\mathcal{P}^{34}+\mathcal{P}^{32}\mathcal{P}%
^{34})\right.  \notag \\
&&\qquad \qquad \qquad \left. +\frac{1}{2}(\mathcal{P}^{13}\mathcal{P}^{42}+%
\mathcal{P}^{13}\mathcal{P}^{43}+\mathcal{P}^{42}\mathcal{P}^{43})\right] .
\end{eqnarray}%
Even if we restrict the sum between the square brackets to the support of
the $\delta $ function, we can easily check that the expression does not
have a proper diagrammatic form, as in the first line of (\ref{reinste}),
because $\mathcal{P}^{34}$ and $\mathcal{P}^{43}$ are unaffected by $\Delta
^{12}$. We can adjust $w_{12}$ by adding and subtracting%
\begin{equation}
\tilde{w}_{12}\equiv -\Delta ^{12}\left[ \frac{1}{2}\left( \mathcal{P}^{14}%
\mathcal{P}^{32}-\mathcal{P}^{14}\mathcal{P}^{34}-\mathcal{P}^{32}\mathcal{P}%
^{34}\right) +\frac{1}{2}\left( \mathcal{P}^{13}\mathcal{P}^{42}-\mathcal{P}%
^{13}\mathcal{P}^{43}-\mathcal{P}^{42}\mathcal{P}^{43}\right) \right] .
\end{equation}%
Indeed, the sum 
\begin{eqnarray}
w_{12}+\tilde{w}_{12} &=&-\Delta ^{12}\left( \mathcal{P}^{13}\mathcal{P}%
^{14}+\mathcal{P}^{32}\mathcal{P}^{42}+\mathcal{P}^{14}\mathcal{P}^{32}+%
\mathcal{P}^{13}\mathcal{P}^{42}\right) =-\Delta ^{12}\mathcal{Q}^{13}%
\mathcal{Q}^{14}  \notag \\
&=&-32\pi ^{2}\omega _{1}\omega _{2}\omega _{3}\omega _{4}\mathcal{P}\int 
\frac{\mathrm{d}k^{0}}{2\pi }\frac{\delta ^{-}((k-p_{1})^{2}-m_{1}^{2})}{%
(k-p_{3})^{2}-m_{3}^{2}}\frac{\delta ^{+}((k-p_{2})^{2}-m_{2}^{2})}{%
(k-p_{4})^{2}-m_{4}^{2}},
\end{eqnarray}%
has a proper diagrammatic form. On the other hand, it is easy to show, by
means of the first identity of formula (\ref{idprin}), that the contribution
we need to subtract is a triple threshold:%
\begin{equation}
\tilde{w}_{12}=\frac{1}{2}\Delta ^{12}\Delta ^{32}\Delta ^{14}+\frac{1}{2}%
\Delta ^{12}\Delta ^{42}\Delta ^{13}.  \label{resi}
\end{equation}%
Therefore, we move it down to level 3.

The terms proportional to the product of two $\delta $ functions in (\ref%
{G4s}), i.e.,%
\begin{equation}
\frac{i}{2}\sum_{\text{perms}}\Delta ^{ab}\Delta ^{ac}\mathcal{P}^{ad}+\frac{%
i}{4}\sum_{\text{perms}}(\Delta ^{ab}\Delta ^{ac}\mathcal{P}^{db}+\mathcal{P}%
^{ab}\Delta ^{ac}\Delta ^{db}+\Delta ^{ab}\mathcal{P}^{ac}\Delta ^{db})
\end{equation}%
plus $(e\rightarrow -e)$, are all double thresholds, since there is no
product of principal values here that needs to be rearranged. For example,
the terms proportional to $\Delta ^{12}\Delta ^{13}$ and $\Delta ^{12}\Delta
^{34}$ (the other possibilities being simple transformations of these),
either vanish or can be readily written in a proper diagrammatic form:%
\begin{equation}
i\Delta ^{12}\Delta ^{13}\left( \mathcal{P}^{14}+\frac{1}{2}\mathcal{P}^{42}+%
\frac{1}{2}\mathcal{P}^{43}\right) =i\Delta ^{12}\Delta ^{13}\mathcal{Q}%
^{14},\qquad \frac{i}{2}\Delta ^{12}\Delta ^{34}\left( \mathcal{P}^{14}+%
\mathcal{P}^{32}\right) =0.
\end{equation}

Finally, the triple thresholds are $\left. G_{\text{ABCD}}^{s4}\right\vert
_{F\rightarrow -i\Delta }$, from formula (\ref{G4s}), plus the contributions
like (\ref{resi}) coming down from the terms proportional to one $\delta $
function.

\subsection{Formulas}

In the end, the threshold decomposition of the box skeleton diagram reads%
\begin{eqnarray}
G_{\text{ABCD}}^{s4} &=&-i\mathcal{P}_{4}-\frac{1}{2}\sum_{\text{perms}%
}\Delta ^{ab}\mathcal{Q}^{ac}\mathcal{Q}^{ad}+\frac{i}{2}\sum_{\text{perms}%
}\Delta ^{ab}(\Delta ^{ac}+\Delta ^{cb})\mathcal{Q}^{ad}  \notag \\
&&\qquad \qquad \qquad \qquad +\frac{1}{6}\sum_{\text{perms}}\Delta
^{ab}(\Delta ^{ac}\Delta ^{ad}+\Delta ^{cb}\Delta ^{db}),  \label{boxfey}
\end{eqnarray}%
where $\mathcal{P}_{4}=i\left. G_{\text{ABCD}}^{s4}\right\vert
_{F\rightarrow \mathcal{P}}$ . We obtain the table 
\begin{equation}
\begin{tabular}{|c|c|c|c|c|c|c|}
\hline
Th$\setminus $G & $G_{\text{ABCD}}^{s4}$ & $G_{\text{\.{A}\.{B}\.{C}\.{D}}%
}^{s4}$ & [ $G_{\text{A\.{B}CD}}^{s4}$ & $G_{\text{\.{A}B\.{C}\.{D}}}^{s4}$
& $G_{\text{\.{A}BC\.{D}}}^{s4}$ & $\frac{1}{2}G_{\text{\.{A}B\.{C}D}}^{s4}$
] \\ \hline
--- & $-i\mathcal{P}_{4}$ & $i\mathcal{P}_{4}$ & $0$ & $0$ & $0$ & $0$ \\ 
\hline
\lbrack\ $\Delta ^{12}$ ] & $-\mathcal{Q}^{13}\mathcal{Q}^{14}$ & $-\mathcal{%
Q}^{13}\mathcal{Q}^{14}$ & $0$ & $2\mathcal{Q}^{13}\mathcal{Q}^{14}$ & $0$ & 
$0$ \\ \hline
\lbrack\ $\Delta ^{21}$ ] & $-\mathcal{Q}^{23}\mathcal{Q}^{24}$ & $-\mathcal{%
Q}^{23}\mathcal{Q}^{24}$ & $2\mathcal{Q}^{23}\mathcal{Q}^{24}$ & $0$ & $0$ & 
$0$ \\ \hline
\lbrack\ $\Delta ^{13}$ ] & $-\mathcal{Q}^{12}\mathcal{Q}^{14}$ & $-\mathcal{%
Q}^{12}\mathcal{Q}^{14}$ & $0$ & $0$ & $2\mathcal{Q}^{12}\mathcal{Q}^{14}$ & 
$0$ \\ \hline
\ldots & \ldots & \ldots & \ldots & \ldots & \ldots & \ldots \\ \hline
\end{tabular}%
\end{equation}%
where we have reported only the single thresholds. The contributions between
brackets must be summed over the cyclic permutations of 1, 2, 3, 4.

The complex conjugate diagram is $G_{\text{\.{A}\.{B}\.{C}\.{D}}}^{s4}=\bar{G%
}_{\text{ABCD}}^{s4}$. The decompositions of the cut diagrams are%
\begin{eqnarray}
G_{\text{\.{A}B\.{C}\.{D}}}^{s4} &=&2\Delta ^{12}\mathcal{Q}^{13}\mathcal{Q}%
^{14}+2i\Delta ^{12}\left[ \Delta ^{13}\mathcal{Q}^{14}+\Delta ^{14}\mathcal{%
Q}^{13}+\Delta ^{32}\mathcal{Q}^{14}+\Delta ^{42}\mathcal{Q}^{13}\right] 
\notag \\
&&-2\Delta ^{12}\left[ \Delta ^{13}\Delta ^{14}+\Delta ^{14}\Delta
^{34}+\Delta ^{13}\Delta ^{43}+\Delta ^{32}\Delta ^{42}\right] ,  \notag \\
G_{\text{\.{A}BC\.{D}}}^{s4} &=&2\Delta ^{13}\mathcal{Q}^{12}\mathcal{Q}%
^{14}+2i\Delta ^{13}\left[ \Delta ^{14}\mathcal{Q}^{12}+\Delta ^{43}\mathcal{%
Q}^{12}-\Delta ^{23}\mathcal{Q}^{14}-\Delta ^{12}\mathcal{Q}^{14}\right] 
\notag \\
&&+2\Delta ^{13}\left[ \Delta ^{12}\Delta ^{14}+\Delta ^{14}\Delta
^{24}+\Delta ^{23}\Delta ^{43}+\Delta ^{42}\Delta ^{43}\right] ,  \notag \\
G_{\text{A\.{B}CD}}^{s4} &=&\left. \bar{G}_{\text{\.{A}B\.{C}\.{D}}%
}^{s4}\right\vert _{e\rightarrow -e},\qquad G_{\text{\.{A}B\.{C}D}%
}^{s4}=8\Delta ^{12}\Delta ^{14}\Delta ^{34},  \label{g4cut}
\end{eqnarray}%
up to cyclic permutations.

Using (\ref{boxfey}) and (\ref{g4cut}), it is easy to check the spectral
optical theorem%
\begin{equation}
G_{\text{ABCD}}^{s4}+G_{\text{\.{A}\.{B}\.{C}\.{D}}}^{s4}+\left[ G_{\text{A%
\.{B}CD}}^{s4}+G_{\text{\.{A}B\.{C}\.{D}}}^{s4}+G_{\text{\.{A}BC\.{D}}}^{s4}+%
\frac{1}{2}G_{\text{\.{A}B\.{C}D}}^{s4}+\text{cycl}\right] =0.  \label{sot4}
\end{equation}%
The factor 1/2 avoids overcounting when the cyclic permutations are
included. As usual, the optical theorem%
\begin{equation}
G_{\text{ABCD}}^{4}+G_{\text{\.{A}\.{B}\.{C}\.{D}}}^{4}+\left[ G_{\text{A\.{B%
}CD}}^{4}+G_{\text{\.{A}B\.{C}\.{D}}}^{4}+G_{\text{\.{A}BC\.{D}}}^{4}+\frac{1%
}{2}G_{\text{\.{A}B\.{C}D}}^{4}+\text{cycl}\right] =0  \label{ot4}
\end{equation}%
is obtained by integrating on the space components of the loop momentum
(after dividing by $2\omega $ for every internal leg). Finally, the spectral
optical identities are obtained by separating the contributions of the
various (multi)thresholds to (\ref{sot4}).

\subsection{Fakeons}

When fakeons are present, we apply the fakeon prescription/projection
mentioned in step 5) of subsection \ref{strat}, which amounts to eliminating
the contributions coming from the thresholds that involve one or more
fakeons. The spectral optical identities and the spectral optical theorem
continue to hold after these operations, since the thresholds do not
interfere with one another.

Concretely, with one fakeon in leg 4, we must suppress all the $\Delta ^{ab}$
with $a$ or $b$ equal to 4. From (\ref{boxfey}), the uncut diagram is given
by%
\begin{equation}
G_{\text{ABCDf}}^{s4}=-i\mathcal{P}_{4}-\sum_{p(1,2,3)}\Delta ^{ab}\mathcal{Q%
}^{ac}\mathcal{Q}^{a4}+\frac{i}{2}\sum_{p(1,2,3)}\Delta ^{ab}(\Delta
^{ac}+\Delta ^{cb})\mathcal{Q}^{a4},
\end{equation}%
where the sum is restricted to the permutations $\{a,b,c\}$ of 1, 2 and 3.
The nonvanishing cut diagrams are%
\begin{eqnarray}
G_{\text{A\.{B}CDf}}^{s4} &=&\left. \bar{G}_{\text{\.{A}B\.{C}\.{D}f}%
}^{s4}\right\vert _{e\rightarrow -e},\qquad G_{\text{\.{A}B\.{C}\.{D}f}%
}^{s4}=2\Delta ^{12}\mathcal{Q}^{14}(\mathcal{Q}^{13}+i\Delta ^{13}+i\Delta
^{32}),  \notag \\
G_{\text{AB\.{C}Df}}^{s4} &=&\left. \bar{G}_{\text{\.{A}\.{B}C\.{D}f}%
}^{s4}\right\vert _{e\rightarrow -e},\qquad G_{\text{\.{A}\.{B}C\.{D}f}%
}^{s4}=2\Delta ^{23}\mathcal{Q}^{24}(\mathcal{Q}^{21}+i\Delta ^{21}+i\Delta
^{13}),  \label{cutBox} \\
G_{\text{\.{A}BC\.{D}f}}^{s4} &=&2\Delta ^{13}\mathcal{Q}^{14}(\mathcal{Q}%
^{12}-i\Delta ^{23}-i\Delta ^{12}),\qquad G_{\text{A\.{B}\.{C}Df}%
}^{s4}=\left. \bar{G}_{\text{\.{A}BC\.{D}f}}^{s4}\right\vert _{e\rightarrow
-e}.  \notag
\end{eqnarray}%
It is easy to verify that these expressions satisfy the spectral optical
theorem 
\begin{equation}
G_{\text{ABCDf}}^{s4}+G_{\text{\.{A}\.{B}\.{C}\.{D}f}}^{s4}+G_{\text{A\.{B}%
CDf}}^{s4}+G_{\text{\.{A}B\.{C}\.{D}f}}^{s4}+G_{\text{AB\.{C}Df}}^{s4}+G_{%
\text{\.{A}\.{B}C\.{D}f}}^{s4}+G_{\text{\.{A}BC\.{D}f}}^{s4}+G_{\text{A\.{B}%
\.{C}Df}}^{s4}=0,
\end{equation}%
where $G_{\text{\.{A}\.{B}\.{C}\.{D}f}}^{s4}=\bar{G}_{\text{ABCDf}}^{s4}$,
and match the diagrammatics of section \ref{diagra}. The vanishing cut
diagrams are those that contain one cut fakeon propagator, which is
identically zero by (\ref{cutva}), so they also agree with the diagrammatics.

With fakeons in 3 and 4 we have%
\begin{equation}
G_{\text{ABCfDf}}^{s4}=-i\mathcal{P}_{4}-\Delta ^{12}\mathcal{Q}^{13}%
\mathcal{Q}^{14}-\Delta ^{21}\mathcal{Q}^{23}\mathcal{Q}^{24},
\end{equation}%
the nonvanishing cut diagrams being%
\begin{equation}
G_{\text{A\.{B}CfDf}}^{s4}=\left. \bar{G}_{\text{\.{A}B\.{C}f\.{D}f}%
}^{s4}\right\vert _{e\rightarrow -e},\qquad G_{\text{\.{A}B\.{C}f\.{D}f}%
}^{s4}=2\Delta ^{12}\mathcal{Q}^{13}\mathcal{Q}^{14}.
\end{equation}%
With fakeons in 2 and 4, we have%
\begin{eqnarray}
G_{\text{ABfCDf}}^{s4} &=&-i\mathcal{P}_{4}-\Delta ^{13}\mathcal{Q}^{12}%
\mathcal{Q}^{14}-\Delta ^{31}\mathcal{Q}^{32}\mathcal{Q}^{34},  \notag \\
G_{\text{A\.{B}f\.{C}Df}}^{s4} &=&\left. \bar{G}_{\text{\.{A}BfC\.{D}f}%
}^{s4}\right\vert _{e\rightarrow -e},\qquad G_{\text{\.{A}BfC\.{D}f}%
}^{s4}=2\Delta ^{13}\mathcal{Q}^{12}\mathcal{Q}^{14}.
\end{eqnarray}%
Finally, with fakeons in three or four legs, we obtain%
\begin{equation}
G_{\text{ABfCfDf}}^{s4}=G_{\text{AfBfCfDf}}^{s4}=-i\mathcal{P}_{4}
\end{equation}%
and no nontrivial cut diagram. This result encodes the purely virtual
content of the box diagram.

\section{Simple multiloop diagrams}

\label{multi}\setcounter{equation}{0} 
\begin{figure}[t]
\begin{center}
\includegraphics[width=14truecm]{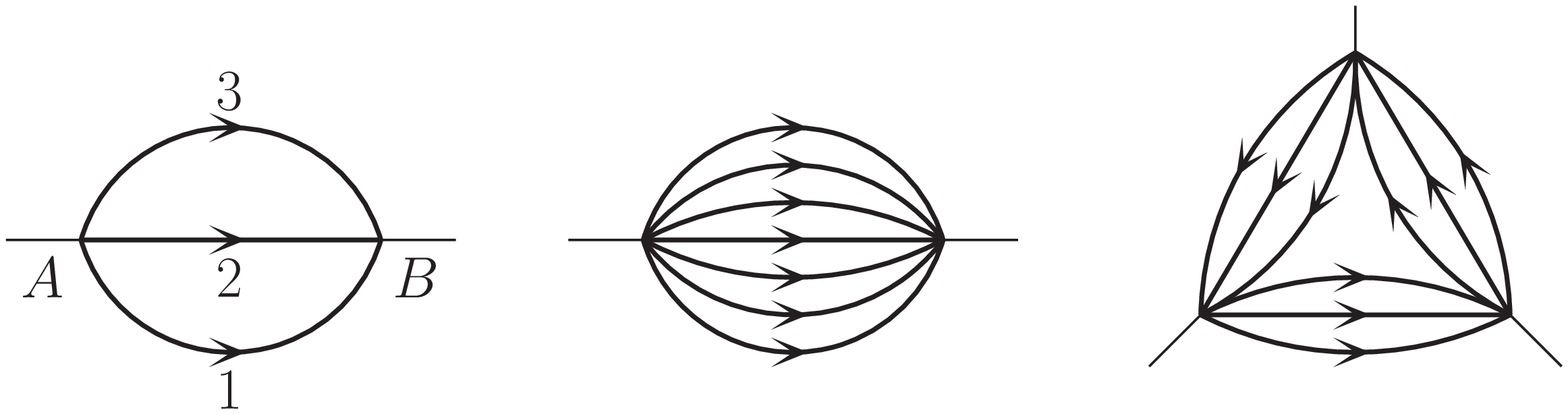}
\end{center}
\caption{Multiloop diagrams with stacks of internal legs connecting vertices}
\label{Fasci}
\end{figure}

In this section we discuss a class of multiloop diagrams that can be treated
straightforwardly by generalizing the techniques applied in the previous
sections. They are those with stacks of internal legs connecting the same
vertices, as shown in Fig. \ref{Fasci}.

Formula (\ref{tb0}) shows that, after the integral on the loop energy, the
single bubble is equivalent\footnote{%
Apart from a minus sign, which can be easily handled as an overall factor in
the spectral optical theorem.} to a propagator with energy equal to the
total incoming energy and frequency equal to the sum of the frequencies.
This property iterates to arbitrary stacks of propagators. For example,
consider the left two-loop diagram of Fig. \ref{Fasci}, which is the bubble
with \textquotedblleft diagonal\textquotedblright . It leads to the skeleton
integral%
\begin{equation}
G_{\text{AB}}^{s2D}=\int \frac{\mathrm{d}k^{0}}{2\pi }\int \frac{\mathrm{d}%
q^{0}}{2\pi }\prod\limits_{i=1}^{2}\frac{2\omega _{i}}{%
(l_{i}-p_{i})^{2}-m_{i}^{2}+i\epsilon _{i}}\frac{2\omega _{3}}{%
(q+k+p_{3})^{2}-m_{3}^{2}+i\epsilon _{3}},
\end{equation}%
where $l_{1}=k$, $l_{2}=q$. Using the residue theorem, we obtain%
\begin{equation}
G_{\text{AB}}^{s2D}=\frac{-2i(\omega _{1}+\omega _{2}+\omega _{3})}{%
(e_{1}+e_{2}+e_{3})^{2}-(\omega _{1}+\omega _{2}+\omega _{3})^{2}+i(\epsilon
_{1}+\epsilon _{2}+\epsilon _{3})}.
\end{equation}%
Again, the stack of propagators is equivalent to a single propagator, from
the point of view of the spectral optical theorem, with energy equal to the
total incoming energy and frequency equal to the sum of the frequencies,
apart from an overall minus sign.

\begin{figure}[t]
\begin{center}
\includegraphics[width=14truecm]{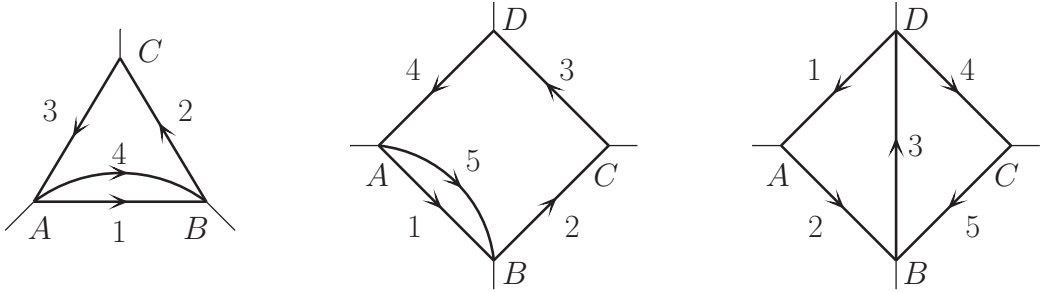}
\end{center}
\caption{Triangle and box diagrams with \textquotedblleft
diagonals\textquotedblright }
\label{BoxD}
\end{figure}

Inside more complicated diagrams, we can replace the stack of propagators
with a single propagator, as just shown, and repeat the analyses of the
previous sections. For example, the triangle of fig. \ref{BoxD} (triangle
with \textquotedblleft diagonal\textquotedblright ) is the triangle of
section \ref{triangleT} with%
\begin{equation}
e_{1}\rightarrow e_{1}+e_{4},\qquad \omega _{1}\rightarrow \omega
_{1}+\omega _{4}.
\end{equation}%
The central diagram of fig. \ref{BoxD} is the box of section \ref{boxG} with%
\begin{equation}
e_{1}\rightarrow e_{1}+e_{5},\qquad \omega _{1}\rightarrow \omega
_{1}+\omega _{5}.
\end{equation}%
It can be used to calculate the left diagram of fig. \ref{2selfB}, which
contributes to the two-loop correction to the self energy. 

As far as the fakeon prescription is concerned, it is sufficient to have one
fakeon leg in the stack to convert the entire stack into a fakeon. The
diagrammatics adapts coherently, since it is impossible to cut a leg
belonging to the stack without cutting the whole stack.

\section{Box with diagonal}

\label{boxwdiag}\setcounter{equation}{0}

\begin{figure}[t]
\begin{center}
\includegraphics[width=6truecm]{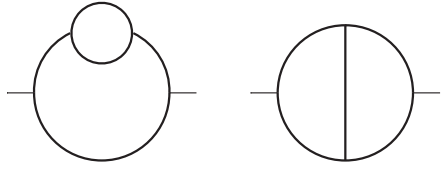}
\end{center}
\caption{Two-loop corrections to the self energy}
\label{2selfB}
\end{figure}

A more interesting two-loop diagram is the box with one diagonal, which we
denote by $G^{4D}$. Referring to Fig. \ref{BoxD}, it is identified by the
\textquotedblleft word\textquotedblright\ (ABD)(BCD), or, in extended
notation, (A2B3D1)(B5C4D3). Below, we just write ABCD. This diagram can be
used, for example, to evaluate the right diagram of fig. \ref{2selfB}, which
is the second contribution to the two-loop self energy. To cover the most
general case, we attach external momenta to every vertex.

Choosing the $p_{i}$ orientations opposite to the arrows, the skeleton is%
\begin{equation}
G_{\text{ABCD}}^{s4D}=i\int \frac{\mathrm{d}k^{0}}{2\pi }\int \frac{\mathrm{d%
}q^{0}}{2\pi }\frac{2\omega _{3}}{(q+k-p_{3})^{2}-m_{3}^{2}+i\epsilon _{3}}%
\prod\limits_{i=1,2,4,5}\frac{2\omega _{i}}{(l_{i}-p_{i})^{2}-m_{i}^{2}+i%
\epsilon _{i}},
\end{equation}%
where $l_{1}=l_{2}=k$, $l_{4}=l_{5}=q$. As usual, we first integrate on the
loop energies by means of the residue theorem. Then, we use the smart
common-denominator identities (\ref{smartid}) to eliminate the
pseudothresholds in favor of the physical thresholds. The result is%
\begin{equation}
G_{\text{ABCD}}^{s4D}=-i\sum_{s4D}F^{a3c}\left[ F^{ab}F^{cd}+\frac{1}{2}%
F^{ab}F^{a3d}+\frac{1}{2}F^{b3c}F^{cd}+\frac{1}{2}F^{a3d}F^{b3c}\right] ,
\end{equation}%
where%
\begin{equation}
F^{abc}=\frac{1}{e_{a}-e_{b}+e_{c}-\tilde{\omega}_{a}-\tilde{\omega}_{b}-%
\tilde{\omega}_{c}}
\end{equation}%
and the sum $\sum_{s4D}$ is on $\{a,b\}=p(1,2)$, $\{c,d\}=p(4,5)$, where $%
p(1,2)$, $p(4,5)$ are the permutations of 1,2 and 4,5, respectively, plus $%
(e\rightarrow -e)$.

To write the threshold decomposition, it is convenient to define%
\begin{eqnarray}
Q^{abc} &=&\mathcal{P}\left( \frac{1}{e_{a}-e_{b}+e_{c}-\omega _{a}-\omega
_{b}-\omega _{c}}-\frac{1}{e_{a}-e_{b}+e_{c}-\omega _{a}+\omega _{b}-\omega
_{c}}\right) ,  \notag \\
\Delta ^{abc} &=&\pi \delta (e_{a}-e_{b}+e_{c}-\omega _{a}-\omega
_{b}-\omega _{c}),\qquad \hat{\Delta}^{abc}=\pi \delta
(e_{a}-e_{b}+e_{c}+\omega _{a}+\omega _{b}+\omega _{c}).\qquad
\end{eqnarray}

As usual, the threshold decomposition is worked out by using (\ref{sgnid})
and writing the various contributions in manifest diagrammatic forms by
means of identities like (\ref{idprin}). The result is%
\begin{eqnarray}
G_{\text{ABCD}}^{s4D} &=&-i\mathcal{P}_{\text{ABCD}}^{s4D}-\Delta
^{12}\left. \mathcal{P}_{345}\right\vert _{e_{3}\rightarrow
e_{3}-e_{2}-\omega _{2}}-\Delta ^{45}\left. \mathcal{P}_{123}\right\vert
_{e_{3}\rightarrow e_{3}-e_{5}-\omega _{5}}-\Delta ^{21}\left. \mathcal{P}%
_{345}\right\vert _{e_{3}\rightarrow e_{3}-e_{1}-\omega _{1}}  \notag \\
&&-\Delta ^{54}\left. \mathcal{P}_{123}\right\vert _{e_{3}\rightarrow
e_{3}-e_{4}-\omega _{4}}-\sum_{s4D}\Delta ^{a3c}\mathcal{Q}^{ab}\mathcal{Q}%
^{cd}+i\sum_{s4D}\Delta ^{a3c}(\Delta ^{ab}\mathcal{Q}^{cd}+\mathcal{Q}%
^{ab}\Delta ^{cd})  \notag \\
&&+\frac{i}{2}\sum_{s4D}\left[ \mathcal{Q}^{a3c}\Delta ^{ab}\Delta
^{cd}+\Delta ^{a3c}(\mathcal{Q}^{ab}\Delta ^{a3d}+\Delta ^{b3c}\mathcal{Q}%
^{cd})\right]  \notag \\
&&+\sum_{s4D}\Delta ^{a3c}\left[ \Delta ^{ab}\Delta ^{cd}+\frac{1}{2}\Delta
^{ab}\Delta ^{a3d}+\frac{1}{2}\Delta ^{b3c}\Delta ^{cd}\right] ,
\end{eqnarray}%
where $\mathcal{P}_{\text{ABCD}}^{s4D}=i\left. G_{\text{ABCD}%
}^{s4D}\right\vert _{F\rightarrow \mathcal{P}}$ and $-i\mathcal{P}_{123}$, $%
-i\mathcal{P}_{345}$\ are the purely virtual contents of the triangle
diagrams with legs 1,2,3 and 3,4,5, respectively (see section \ref{triangleT}%
, formula (\ref{Tdecomp})). The decompositions of the cut diagrams are%
\begin{eqnarray}
G_{\text{\.{A}BCD}}^{s4D} &=&2i\Delta ^{21}\left. T_{345}^{s\hspace{0.01in}%
}\right\vert _{e_{3}\rightarrow e_{3}-e_{1}-\omega _{1}},\quad G_{\text{A\.{B%
}\.{C}\.{D}}}^{s4D}=\left. \bar{G}_{\text{\.{A}BCD}}^{s4D}\right\vert
_{e\rightarrow -e},\quad G_{\text{AB\.{C}D}}^{s4D}=\left. G_{\text{\.{A}BCD}%
}^{s4D}\right\vert _{2\leftrightarrow 5}^{1\leftrightarrow 4},  \notag \\
G_{\text{\.{A}\.{B}C\.{D}}}^{s4D} &=&\left. G_{\text{A\.{B}\.{C}\.{D}}%
}^{s4D}\right\vert _{2\leftrightarrow 5}^{1\leftrightarrow 4},\qquad G_{%
\text{ABC\.{D}}}^{s4D}=\left. \bar{G}_{\text{\.{A}B\.{C}\.{D}}%
}^{s4D}\right\vert _{4\leftrightarrow 5}^{1\leftrightarrow 2},\qquad G_{%
\text{\.{A}\.{B}\.{C}D}}^{s4D}=\left. \bar{G}_{\text{A\.{B}CD}%
}^{s4D}\right\vert _{4\leftrightarrow 5}^{1\leftrightarrow 2},  \notag \\
G_{\text{A\.{B}CD}}^{s4D} &=&2\hat{\Delta}^{235}(\mathcal{\hat{Q}}%
^{21}-i\Delta ^{12}-i\hat{\Delta}^{135})(\mathcal{\hat{Q}}^{54}-i\Delta
^{45}-i\hat{\Delta}^{234}),\qquad G_{\text{\.{A}B\.{C}\.{D}}}^{s4D}=\left. 
\bar{G}_{\text{A\.{B}CD}}^{s4D}\right\vert _{e\rightarrow -e},  \notag \\
G_{\text{\.{A}\.{B}CD}}^{s4D} &=&2\hat{\Delta}^{135}(\mathcal{\hat{Q}}%
^{12}+i\Delta ^{21}+i\hat{\Delta}^{235})(\mathcal{\hat{Q}}^{54}-i\Delta
^{45}-i\hat{\Delta}^{134}),\qquad G_{\text{AB\.{C}\.{D}}}^{s4D}=\left. \bar{G%
}_{\text{\.{A}\.{B}CD}}^{s4D}\right\vert _{e\rightarrow -e},  \notag \\
G_{\text{A\.{B}\.{C}D}}^{s4D} &=&2\hat{\Delta}^{234}(\mathcal{\hat{Q}}%
^{21}-i\Delta ^{12}-i\hat{\Delta}^{134})(\mathcal{\hat{Q}}^{45}+i\Delta
^{54}+i\hat{\Delta}^{235}),\qquad G_{\text{\.{A}BC\.{D}}}^{s4D}=\left. \bar{G%
}_{\text{A\.{B}\.{C}D}}^{s4D}\right\vert _{e\rightarrow -e},  \notag \\
G_{\text{\.{A}B\.{C}D}}^{s4D} &=&4i\Delta ^{21}\Delta ^{54}(\mathcal{Q}%
^{235}-i\Delta ^{235}-i\hat{\Delta}^{134}),\qquad G_{\text{A\.{B}C\.{D}}%
}^{s4D}=\left. \bar{G}_{\text{\.{A}B\.{C}D}}^{s4D}\right\vert _{e\rightarrow
-e},
\end{eqnarray}%
where $T_{345}^{s\hspace{0.01in}}$\ is the triangle (\ref{Tdecomp}) with
legs 3,4,5. The conjugate uncut diagram is $G_{\text{\.{A}\.{B}\.{C}\.{D}}%
}^{s4D}=\bar{G}_{\text{ABCD}}^{s4D}$.

The spectral optical theorem reads 
\begin{eqnarray}
&&G_{\text{ABCD}}^{s4D}+G_{\text{\.{A}\.{B}\.{C}\.{D}}}^{s4D}+G_{\text{\.{A}%
BCD}}^{s4D}+G_{\text{A\.{B}\.{C}\.{D}}}^{s4D}+G_{\text{AB\.{C}D}}^{s4D}+G_{%
\text{\.{A}\.{B}C\.{D}}}^{s4D}+G_{\text{ABC\.{D}}}^{s4D}+G_{\text{\.{A}\.{B}%
\.{C}D}}^{s4D}+G_{\text{A\.{B}CD}}^{s4D}  \notag \\
&&\phantom{C}+G_{\text{\.{A}B\.{C}\.{D}}}^{s4D}+G_{\text{\.{A}\.{B}CD}%
}^{s4D}+G_{\text{AB\.{C}\.{D}}}^{s4D}+G_{\text{A\.{B}\.{C}D}}^{s4D}+G_{\text{%
\.{A}BC\.{D}}}^{s4D}+G_{\text{\.{A}B\.{C}D}}^{s4D}+G_{\text{A\.{B}C\.{D}}%
}^{s4D}=0
\end{eqnarray}%
and can be easily verified. The spectral optical identities are the various
threshold contributions to this equality, which vanish separately.

\subsection{Fakeons}

The fakeon projections can now be implemented straightforwardly. If leg 1,
which is the segment DA, is a fakeon, we suppress all the contributions
proportional to $\Delta ^{ab}$, $\Delta ^{abc}$ and $\hat{\Delta}^{abc}$,
whenever $a$, $b$ or $c$ are equal to 1. Denoting this case by fABCD, the
uncut diagram is%
\begin{eqnarray}
G_{\text{fABCD}}^{s4D} &=&-i\mathcal{P}_{\text{ABCD}}^{s4D}-\Delta
^{45}(\left. \mathcal{P}_{123}\right\vert _{e_{3}\rightarrow
e_{3}-e_{5}-\omega _{5}}-i\Delta ^{234}\mathcal{Q}^{21})-\Delta ^{234}%
\mathcal{Q}^{21}(\mathcal{Q}^{45}-i\Delta ^{235})  \notag \\
&&-\Delta ^{54}\left. (\mathcal{P}_{123}\right\vert _{e_{3}\rightarrow
e_{3}-e_{4}-\omega _{4}}-i\hat{\Delta}^{234}\mathcal{\hat{Q}}^{21})-\hat{%
\Delta}^{234}\mathcal{\hat{Q}}^{21}(\mathcal{\hat{Q}}^{45}-i\hat{\Delta}%
^{235})  \notag \\
&&-\Delta ^{235}\mathcal{Q}^{21}(\mathcal{Q}^{54}-i\Delta ^{54})-\hat{\Delta}%
^{235}\mathcal{\hat{Q}}^{21}(\mathcal{\hat{Q}}^{54}-i\Delta ^{45}).
\end{eqnarray}

The cut diagrams $G_{\text{f\.{A}BCD}}^{s4D}$, $G_{\text{fA\.{B}\.{C}\.{D}}%
}^{s4D}$, $G_{\text{fABC\.{D}}}^{s4D}$, $G_{\text{f\.{A}\.{B}\.{C}D}}^{s4D}$%
, $G_{\text{f\.{A}\.{B}CD}}^{s4D}$, $G_{\text{fAB\.{C}\.{D}}}^{s4D}$, $G_{%
\text{f\.{A}B\.{C}D}}^{s4D}$ and $G_{\text{fA\.{B}C\.{D}}}^{s4D}$ disappear
altogether. This is consistent with the diagrammatics of the fakeon
prescription, since those diagrams contain a cut fakeon leg and by (\ref%
{cutva}) the cut fakeon propagator vanishes. The surviving cut diagrams are%
\begin{eqnarray}
G_{\text{fAB\.{C}D}}^{s4D} &=&2\Delta ^{54}(\left. \mathcal{P}%
_{123}\right\vert _{e_{3}\rightarrow e_{3}-e_{4}-\omega _{4}}-i\Delta ^{235}%
\mathcal{Q}^{21}-i\hat{\Delta}^{234}\mathcal{\hat{Q}}^{21}),  \notag \\
G_{\text{f\.{A}\.{B}C\.{D}}}^{s4D} &=&2\Delta ^{45}(\left. \mathcal{P}%
_{123}\right\vert _{e_{3}\rightarrow e_{3}-e_{5}-\omega _{5}}+i\Delta ^{234}%
\mathcal{Q}^{21}+i\hat{\Delta}^{235}\mathcal{\hat{Q}}^{21}),  \notag \\
G_{\text{fA\.{B}CD}}^{s4D} &=&2\hat{\Delta}^{235}\mathcal{\hat{Q}}^{21}(%
\mathcal{\hat{Q}}^{54}-i\Delta ^{45}-i\hat{\Delta}^{234}),\quad G_{\text{f\.{%
A}B\.{C}\.{D}}}^{s4D}=2\Delta ^{235}\mathcal{Q}^{21}(\mathcal{Q}%
^{54}+i\Delta ^{54}+i\Delta ^{234}),  \notag \\
G_{\text{fA\.{B}\.{C}D}}^{s4D} &=&2\hat{\Delta}^{234}\mathcal{\hat{Q}}^{21}(%
\mathcal{\hat{Q}}^{45}+i\Delta ^{54}+i\hat{\Delta}^{235}),\quad G_{\text{f\.{%
A}BC\.{D}}}^{s4D}=2\Delta ^{234}\mathcal{Q}^{21}(\mathcal{Q}^{45}-i\Delta
^{45}-i\Delta ^{235}),\qquad  \label{cut4D}
\end{eqnarray}%
It is easy to show that these formulas also follow from the diagrammatic
rules given in subsection \ref{diagra}. In particular, the fakeon propagator
of every link agrees with (\ref{link}).

The spectral optical theorem reads%
\begin{equation}
G_{\text{fABCD}}^{s4D}+G_{\text{f\.{A}\.{B}\.{C}\.{D}}}^{s4D}+G_{\text{fAB%
\.{C}D}}^{s4D}+G_{\text{f\.{A}\.{B}C\.{D}}}^{s4D}+G_{\text{fA\.{B}CD}%
}^{s4D}+G_{\text{f\.{A}B\.{C}\.{D}}}^{s4D}+G_{\text{fA\.{B}\.{C}D}}^{s4D}+G_{%
\text{f\.{A}BC\.{D}}}^{s4D}=0,
\end{equation}%
where $G_{\text{f\.{A}\.{B}\.{C}\.{D}}}^{s4D}=\bar{G}_{\text{fABCD}}^{s4D}$,
and can be verified straightforwardly.

\bigskip

If leg 3 is a fakeon, we denote the diagram by AB$|$f$|$CD. We have%
\begin{eqnarray}
G_{\text{AB$|$f$|$CD}}^{s4D} &=&-i\mathcal{P}_{\text{ABCD}}^{s4D}-\Delta
^{12}\left. \mathcal{P}_{345}\right\vert _{e_{3}\rightarrow
e_{3}-e_{2}-\omega _{2}}-\Delta ^{45}\left. \mathcal{P}_{123}\right\vert
_{e_{3}\rightarrow e_{3}-e_{5}-\omega _{5}}  \notag \\
&&-\Delta ^{21}\left. \mathcal{P}_{345}\right\vert _{e_{3}\rightarrow
e_{3}-e_{1}-\omega _{1}}-\Delta ^{54}\left. \mathcal{P}_{123}\right\vert
_{e_{3}\rightarrow e_{3}-e_{4}-\omega _{4}}+\frac{i}{2}\sum_{s4D}\mathcal{Q}%
^{a3c}\Delta ^{ab}\Delta ^{cd}\qquad 
\end{eqnarray}%
and $G_{\text{\.{A}\.{B}$|$f$|$\.{C}\.{D}}}^{s4D}=\bar{G}_{\text{AB$|$f$|$CD}%
}^{s4D}$. The nonvanishing cut diagrams are%
\begin{eqnarray}
G_{\text{\.{A}B$|$f$|$CD}}^{s4D} &=&2\Delta ^{21}(\left. \mathcal{P}%
_{345}\right\vert _{e_{3}\rightarrow e_{3}-e_{1}-\omega _{1}}-i\Delta ^{45}%
\mathcal{Q}^{234}-i\Delta ^{54}\mathcal{Q}^{235}),\quad   \notag \\
G_{\text{A\.{B}$|$f$|$\.{C}\.{D}}}^{s4D} &=&2\Delta ^{12}(\left. \mathcal{P}%
_{345}\right\vert _{e_{3}\rightarrow e_{3}-e_{2}-\omega _{2}}+i\Delta ^{54}%
\mathcal{Q}^{135}+i\Delta ^{45}\mathcal{Q}^{134}),\qquad G_{\text{\.{A}B$|$f$%
|$\.{C}D}}^{s4D}=4i\Delta ^{21}\Delta ^{54}\mathcal{Q}^{235},  \notag \\
G_{\text{AB$|$f$|$\.{C}D}}^{s4D} &=&2\Delta ^{54}(\left. \mathcal{P}%
_{123}\right\vert _{e_{3}\rightarrow e_{3}-e_{4}-\omega _{4}}-i\Delta ^{12}%
\mathcal{Q}^{135}-i\Delta ^{21}\mathcal{Q}^{235}),\qquad G_{\text{A\.{B}$|$f$%
|$C\.{D}}}^{s4D}=-4i\Delta ^{12}\Delta ^{45}\mathcal{Q}^{134},  \notag \\
G_{\text{\.{A}\.{B}$|$f$|$C\.{D}}}^{s4D} &=&2\Delta ^{45}(\left. \mathcal{P}%
_{123}\right\vert _{e_{3}\rightarrow e_{3}-e_{5}-\omega _{5}}+i\Delta ^{21}%
\mathcal{Q}^{234}+i\Delta ^{12}\mathcal{Q}^{134}),
\end{eqnarray}%
which agree with the diagrammatic rules of the fakeon prescription.

The spectral optical theorem reads%
\begin{equation}
G_{\text{AB$|$f$|$CD}}^{s4D}+G_{\text{\.{A}\.{B}$|$f$|$\.{C}\.{D}}}^{s4D}+G_{%
\text{\.{A}B$|$f$|$CD}}^{s4D}+G_{\text{A\.{B}$|$f$|$\.{C}\.{D}}}^{s4D}+G_{%
\text{AB$|$f$|$\.{C}D}}^{s4D}+G_{\text{\.{A}\.{B}$|$f$|$C\.{D}}}^{s4D}+G_{%
\text{\.{A}B$|$f$|$\.{C}D}}^{s4D} +G_{\text{A\.{B}$|$f$|$C\.{D}}}^{s4D}=0.
\end{equation}

\bigskip

The other possibilities of distributing fakeons in the internal legs can be
treated similarly. The case where all the internal legs are fakeons gives
the purely virtual content of the box with diagonal, which is%
\begin{equation}
\left. G_{\text{ABCD}}^{s4D}\right\vert _{\text{purely virtual}}=-i\mathcal{P%
}_{\text{ABCD}}^{s4D}.
\end{equation}

\section{Further insight into the algebraic structure of the spectral
optical identities}

\label{smartsec}\setcounter{equation}{0}

In this section we provide more insight into the algebraic structure of the
spectral optical identities and work out formulas for more complicated
diagrams, like the pentagon and the hexagon.

It is useful to introduce a few definitions, such as%
\begin{eqnarray}
I_{N} &=&\sum_{p_{N}}\prod\limits_{k=2}^{N}F^{a_{1}a_{k}}+(e\rightarrow
-e),\qquad
J_{N}=\sum_{p_{N}}F^{a_{N}a_{N-1}}\prod\limits_{k=2}^{N-1}F^{a_{1}a_{k}}+(e%
\rightarrow -e),  \notag \\
K_{N}
&=&\sum_{p_{N}}F^{a_{1}a_{2}}F^{a_{3}a_{2}}F^{a_{3}a_{4}}F^{a_{5}a_{4}}%
\cdots +(e\rightarrow -e),
\end{eqnarray}%
where $p_{N}$ denotes the permutations of $\{a_{1}\cdots a_{N}\}$. 
\begin{figure}[t]
\begin{center}
\includegraphics[width=14truecm]{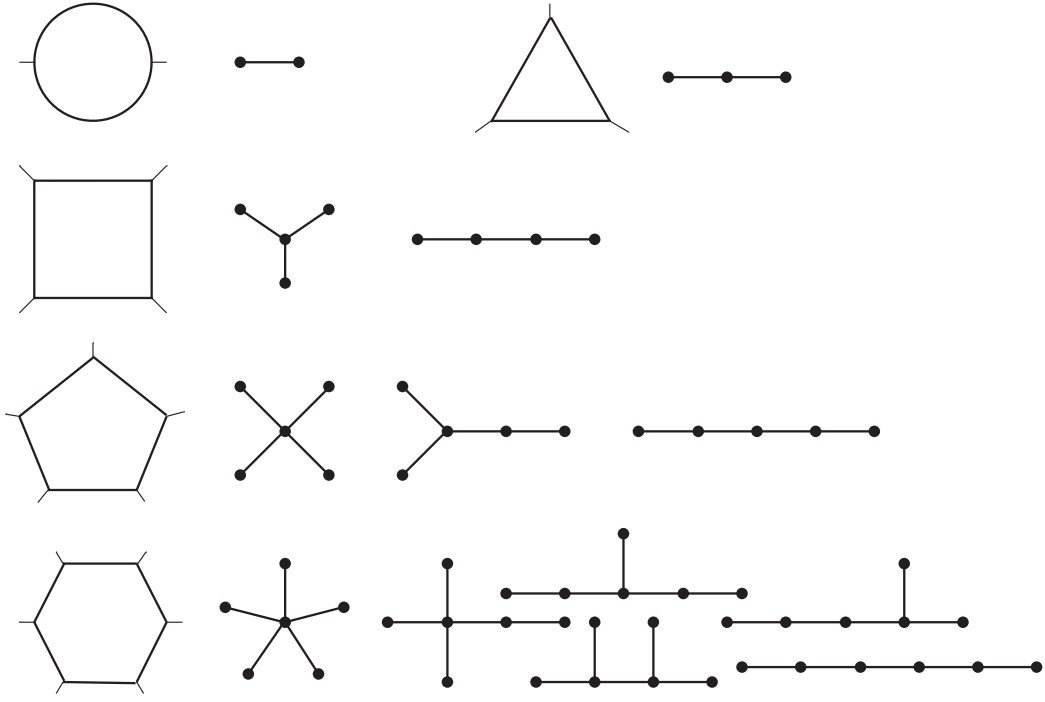}
\end{center}
\caption{One-loop diagrams and their snowflake versions}
\label{oneloopfig}
\end{figure}

The triangle, box, pentagon and hexagon diagrams of fig. \ref{oneloopfig}
give%
\begin{eqnarray}
G_{3}^{s} &=&-\frac{i}{2!}I_{3},\qquad G_{4}^{s}=-\frac{i}{3!}\left( I_{4}+%
\frac{3}{2}J_{4}\right) ,\qquad G_{5}^{s}=-\frac{i}{5!}\left(
I_{5}+4J_{5}+2K_{5}\right) ,  \notag \\
G_{6}^{s} &=&-\frac{i}{6!}\left( I_{6}+5J_{6}+5X_{6}+10Y_{6}\right) ,
\label{boxa}
\end{eqnarray}%
where%
\begin{eqnarray}
X_{6} &=&\sum_{p_{N}}F^{ab}F^{ac}F^{ad}F^{ed}F^{ef}+(e\rightarrow -e),\qquad
Y_{6}=\sum_{p_{N}}F^{ab}F^{ac}F^{ad}F^{ed}F^{fc}+(e\rightarrow -e),  \notag
\\
Z_{6} &=&\sum_{p_{N}}F^{ab}F^{ac}F^{ad}F^{ed}F^{fd}+(e\rightarrow -e),\qquad
Z_{6}+K_{6}=2Y_{6}.  \label{xyz}
\end{eqnarray}

The various contributions can be represented by means of the
\textquotedblleft snowflake diagrams\textquotedblright\ shown in fig. \ref%
{oneloopfig}, where each leg is an $F^{ab}$ and the vertices are repeated
indices. Oriented legs can be used to distinguish $F^{ab}$ from $F^{ba}$.

In general, the multithresholds can be expressed in more than one equivalent
ways. This fact leads to nontrivial identities, such as the last of (\ref%
{xyz}) and%
\begin{equation}
F^{12}F^{13}F^{43}+F^{12}F^{42}F^{43}-F^{12}F^{13}F^{42}-F^{13}F^{42}F^{43}=0.
\label{ambig}
\end{equation}%
Now we show that the threshold decomposition and the fakeon projection are
not affected by this.

Step 4.b) of subsection \ref{strat} tells us how to arrange the levels $\ell
\geqslant 1$ of the decomposition. The proper diagrammatic structure of the $%
\ell =0$ sector, instead, is defined by the procedure itself. The threshold
decomposition of (\ref{ambig}) gives the second identity of (\ref{idprin})
at level 0, that is to say,%
\begin{equation}
\mathcal{P}^{12}\mathcal{P}^{13}\mathcal{P}^{43}+\mathcal{P}^{12}\mathcal{P}%
^{42}\mathcal{P}^{43}-\mathcal{P}^{12}\mathcal{P}^{13}\mathcal{P}^{42}-%
\mathcal{P}^{13}\mathcal{P}^{42}\mathcal{P}^{43}=0.  \label{PPP4}
\end{equation}%
If we add the left-hand side to the purely virtual part of the box diagram,
we change its form, but not its value, because the right-hand side of (\ref%
{PPP4}) vanishes identically. For the same reason, no remnants drop to the
higher levels. Similar arguments apply to the identity of (\ref{xyz}) for
the hexagon diagram.

A source of worry could come from the other identities of (\ref{idprin}),
where the right-hand side is nonvanishing and different levels can mix.
Those identities, however, require an odd number of internal legs (triangle,
pentagon, etc.), i.e., products of an even number of $F^{ab}$. This makes
relations like (\ref{ambig}) unavailable: it is impossible to arrange the
product of an even number of $F^{ab}$ without involving pseudothresholds,
which have disappeared after step 3) of subsection \ref{strat}. In the end,
the threshold decomposition is unambiguous.

We conclude by discussing the algebraic structure of the cut diagrams. In
the case of the triangle, the key identities are%
\begin{equation}
\alpha \beta -\bar{\alpha}\bar{\beta}=(\alpha -\bar{\alpha})\beta +\bar{%
\alpha}(\beta -\bar{\beta})=\alpha (\beta -\bar{\beta})+(\alpha -\bar{\alpha}%
)\bar{\beta},  \label{idopt}
\end{equation}%
where $\alpha $ and $\beta $ are arbitrary complex numbers and $\bar{\alpha}$%
, $\bar{\beta}$ are their complex conjugates. Choosing, e.g., $\alpha
=F^{12} $ and $\beta =F^{13}$ and multiplying by a further factor $-i$, we
get, from the first equality,%
\begin{equation}
-iF^{12}F^{13}+i\bar{F}^{12}\bar{F}^{13}=-2\Delta ^{12}F^{13}-2\bar{F}%
^{12}\Delta ^{13}.
\end{equation}%
The differences $\alpha -\bar{\alpha}$ and $\beta -\bar{\beta}$ give the $%
\delta $ functions belonging to the cut propagators. The application of
identities like these to (\ref{ts3}) gives a different procedure to obtain
the spectral optical identities of the triangle, collected in table \ref{ts1}%
.

In the case of the box, we can manipulate (\ref{G4s}) by means of identities
like%
\begin{eqnarray}
\alpha \beta \gamma -\bar{\alpha}\bar{\beta}\bar{\gamma} &=&(\alpha -\bar{%
\alpha})\beta \gamma +\bar{\alpha}(\beta -\bar{\beta})\gamma +\bar{\alpha}%
\bar{\beta}(\gamma -\bar{\gamma})  \notag \\
&=&\alpha \beta (\gamma -\bar{\gamma})+\alpha (\beta -\bar{\beta})\bar{\gamma%
}+(\alpha -\bar{\alpha})\bar{\beta}\bar{\gamma}  \notag \\
&=&\alpha (\beta -\bar{\beta})\gamma +(\alpha -\bar{\alpha})\bar{\beta}%
\gamma +\bar{\alpha}\bar{\beta}(\gamma -\bar{\gamma})  \notag \\
&=&(\alpha -\bar{\alpha})\bar{\beta}\gamma +\alpha (\beta -\bar{\beta})\bar{%
\gamma}+\bar{\alpha}\beta (\gamma -\bar{\gamma})+(\alpha -\bar{\alpha}%
)(\beta -\bar{\beta})(\gamma -\bar{\gamma})\qquad  \label{idopta}
\end{eqnarray}%
and then apply the threshold decomposition. Note that the last identity
involves cut diagrams with two marked and two unmarked vertices (such as $G_{%
\text{\.{A}BC\.{D}}}^{s4}$).

We can proceed similarly for the pentagon and the hexagon, as well as for
the multiloop diagrams. Identities like (\ref{idopt}) and (\ref{idopta}) can
be used to provide alternative proofs of the optical theorem (via its
spectral version and the spectral optical identities).

In the end, the algebraic structure of the spectral optical identities is
encoded in simple relations such as (\ref{idopt}) and (\ref{idopta}). The
counterpart of this simplicity is a lengthier diagrammatics, in the sense
that each ordinary diagram is expanded into the sum of numerous
\textquotedblleft snowflake\textquotedblright\ diagrams, as shown in fig. %
\ref{oneloopfig}.

\section{Generalization to propagators with arbitrary real residues}

\label{arbres}\setcounter{equation}{0}

In this section we generalize the spectral optical identities to propagators
with arbitrary real residues at the poles. This allows us to treat particles
and antiparticles asymmetrically and shows that the identities do not rely
on Lorentz invariance, nor the CPT theorem.

Define the propagators%
\begin{eqnarray}
\phantom{\bullet}%
\raisebox{-2pt}{\resizebox{2cm}{!}{$\overset{p\rightarrow
}{\!{\raisebox{2.5pt}{\rule{33pt}{.6pt}}}}$}}\!{\phantom{\bullet}} &=&\frac{%
i\xi }{e-\omega +i\epsilon }-\frac{i\zeta }{e+\omega -i\epsilon },  \notag \\
\bullet 
\raisebox{-2pt}{\resizebox{2cm}{!}{$\overset{p\rightarrow }{
\!{\raisebox{2.5pt}{\rule{33pt}{.6pt}}}}$}}\!{\bullet } &=&\frac{i\zeta }{%
e+\omega +i\epsilon }-\frac{i\xi }{e-\omega -i\epsilon },  \notag \\
\bullet 
\raisebox{-2pt}{\resizebox{2cm}{!}{$\overset{p\rightarrow }{
\!{\raisebox{2.5pt}{\rule{33pt}{.6pt}}}}$}}\!{\phantom{\bullet}} &=&(2\pi
)\zeta \delta (e+\omega ),\qquad \bullet \!%
\raisebox{-2pt}{\resizebox{2cm}{!}{$\overset{\leftarrow
p}{{\raisebox{2.5pt}{\rule{33pt}{.6pt}}}}$}}\!{\phantom{\bullet}}=(2\pi )\xi
\delta (e-\omega ).  \label{propagarbi}
\end{eqnarray}%
instead of (\ref{propag}), where $\xi $ and $\zeta $ are arbitrary real
numbers.

For the threshold decompositions, it is convenient to introduce the
quantities 
\begin{eqnarray}
F^{ab} &=&\frac{\xi _{b}}{e_{a}-e_{b}-\tilde{\omega}_{a}-\tilde{\omega}_{b}}%
,\qquad \mathcal{P}^{ab}=\mathcal{P}\frac{\xi _{b}}{e_{a}-e_{b}-\omega
_{a}-\omega _{b}},  \notag \\
\mathcal{Q}^{ab} &=&\mathcal{P}^{ab}-\mathcal{P}\frac{\zeta _{b}}{%
e_{a}-e_{b}-\omega _{a}+\omega _{b}},\qquad \Delta ^{ab}=\pi \zeta _{a}\xi
_{b}\delta (e_{a}-e_{b}-\omega _{a}-\omega _{b}),
\end{eqnarray}%
instead of (\ref{defis}). In the formulas of this section it is understood
that every expression must be expanded and, after the expansion, every power
of $\xi _{a}$ with the same index $a$ must be turned into $\xi _{a}$. The
same must be done for the powers of $\zeta _{a}$. For example, $\Delta
^{ab}\Delta ^{ac}$ is a double $\delta $ function multiplied by $\pi
^{2}\zeta _{a}\xi _{b}\xi _{c}$, instead of $\pi ^{2}\zeta _{a}^{2}\xi
_{b}\xi _{c}$. With these conventions, it is easy to show that formula (\ref%
{G4s}) for the box skeleton turns into%
\begin{equation}
G_{\text{ABCD}}^{s4}=-\frac{i}{6}\sum_{\text{perms}}\zeta
_{a}F^{ab}F^{ac}F^{ad}-\frac{i}{4}\sum_{\text{perms}}\zeta _{a}\zeta
_{d}F^{ab}F^{ac}F^{db}+(e\rightarrow -e,\xi \leftrightarrow \zeta ).
\label{G4sarb}
\end{equation}%
Moreover, the threshold decompositions (\ref{boxfey}) and (\ref{g4cut})\
remain formally the same, like all the other formulas of section \ref{boxG},
including those referring to the fakeon prescription.

The example of the box illustrates how to proceed for every diagram, when we
want to treat propagators with arbitrary real residues.

\section{Diagrams with nontrivial numerators and degenerate diagrams}

\label{degenerate}\setcounter{equation}{0}

So far, we have considered diagrams generated by vertices that do not carry
derivatives. Derivative vertices bring nontrivial numerators into the
integrands. The Passarino-Veltman reduction allows us to convert one-loop
diagrams with arbitrary numerators into linear combinations of diagrams with
unit numerators \cite{pass}. However, it does not work for arbitrary
diagrams. In particular, it may fail with more loops.

The versatility of the spectral optical identities provides a more powerful
way out. Recall that every identity holds without integrating on the space
components $\mathbf{k}_{l}$ of the loop momenta. This means that the
functions of $\mathbf{k}_{l}$ can be factored out. A nontrivial numerator in
a loop integral is a tensorial polynomial $N(p,k)$ of the external momenta $%
p $ and the loop momenta $k$. Let us expand it as a sum of monomials%
\begin{equation}
\prod_{j}p_{j}^{\mu _{j}}\prod_{l}k_{l}^{\mu _{l}}.
\end{equation}%
Since we can factor out the components of $p$ and every space components of $%
k$, we just need to pay attention to the loop energies $k_{l}^{0}$. They can
be used to simplify the poles of the propagators (\ref{propag}) and reduce
the integral to a sum of integrals with fewer internal legs and/or fewer
loop energies in the numerators\footnote{%
If the power of some loop energy $k^{0}$ is large enough, remnants with no $%
k $-dependent denominators can survive. These contributions can be dropped.
If the surviving power of $k^{0}$ is odd, they vanish by symmetric
integration. If the surviving power of $k^{0}$ is even, they are killed by
the dimensional regularization, since they give integrals of polynomials of $%
k$ in $\mathrm{d}^{D}k/(2\pi )^{D}$.}.

Inside the integrals obtained this way, we may not find the propagators (\ref%
{propag}) and (\ref{link}), since the poles at $e=\pm \omega $ need not
appear in those combinations. Yet, we can use the generalization of the
previous section, since formulas (\ref{propagarbi}) with appropriate choices
of $\xi $ or $\zeta $ suit every case.

Iterating this procedure, we reduce to a linear combination of spectral
optical identities of the types already considered. This proves that they
hold for derivative vertices and diagrams with arbitrary numerators.

Another important point is that the identities derived so far have are valid
under a tacit \textquotedblleft non-coincidence\textquotedblright\
assumption, which means that the thresholds are all distinct. The simplest
way to fulfill this requirement is to imagine that the propagators have
nonvanishing, different masses $m_{i}$, such that the sums $\sum_{i\in
J}m_{i}$ are all different, for every subset $J$ of internal legs\footnote{%
To prove the optical theorem in the presence of massless fields, it is
convenient to equip them with small, fictitious masses, otherwise the
asymptotic states of other particles (e.g., the electron in QED) are ill
defined. The non coincidence assumption can be fulfilled using the
fictitious masses. The massless limit is studied after taking care of the
infrared divergences (see section \ref{massless}).}. Another possibility is
to assume, as we have done so far, that each vertex carries an external leg,
equipped with an independent external momentum. For example, instead of the
double bubble $\rangle \hspace{-0.18em}{\bigcirc \hspace{-0.14em}{\bigcirc 
\hspace{-0.18em}\langle }}$, where the central vertex is attached to
internal legs only, we take $\rangle \hspace{-0.18em}{\bigcirc \hspace{%
-0.18em}{|\hspace{-0.18em}{\bigcirc \hspace{-0.18em}\langle }}}$, with
additional external legs stemming from the central vertex.

Yet, in many physical situations identical propagators appear, as in the
left diagram of fig. \ref{2selfB}, and identical thresholds. The square of a
propagator defined by the Feynman prescription is well defined. What about
the threshold decomposition and the fakeon prescription? The square of the
Cauchy principal value is ill-defined and so is the square of a Dirac $%
\delta $ function.

The way out is to use a \textquotedblleft coincidence splitting
method\textquotedblright\ (which works for powers of the Feynman propagator
as well). In other words, we view coinciding thresholds as the limits of
distinct thresholds, obtained in the ways described above (i.e., by
inserting fictitious, small mass differences or equipping the vertices with
additional external legs flowing in small momenta). For example, we can view
the left diagram of fig. \ref{2selfB} as a limit of the middle diagram of
fig. \ref{BoxD}.

Then, arbitrary powers of the Cauchy principal value are well defined, since 
\cite{FLRW}%
\begin{equation}
\lim_{\epsilon \rightarrow 0}\mathcal{P}\prod\nolimits_{i=1}^{n+1}\frac{1}{%
x-\epsilon c_{i}}=\frac{(-1)^{n}}{n!}\frac{\mathrm{d}^{n}}{\mathrm{d}x^{n}}%
\mathcal{P}\frac{1}{x},  \label{PPP}
\end{equation}%
where $c_{i}$ are distinct numbers. The coincidence splitting method
trivializes the powers of a $\delta $ function:%
\begin{equation}
\lim_{\epsilon \rightarrow 0}\prod\nolimits_{i=1}^{n+1}\mathcal{\delta (}%
x-\epsilon c_{i})=0.
\end{equation}%
The Feynman prescription generates products of delta functions times
principal values, but only in well-defined combinations such as%
\begin{equation}
\lim_{\epsilon \rightarrow 0}\left( \frac{\mathcal{\delta (}x-\epsilon c_{1})%
}{x-\epsilon c_{2}}+\frac{\mathcal{\delta (}x-\epsilon c_{2})}{x-\epsilon
c_{1}}\right) =\lim_{\epsilon \rightarrow 0}\frac{\mathcal{\delta (}%
x-\epsilon c_{1})-\mathcal{\delta (}x-\epsilon c_{2})}{\epsilon (c_{1}-c_{2})%
}=-\delta ^{\prime }(x).
\end{equation}%
The fakeon prescription leaves or drops both contributions appearing on the
left-hand side, so it is also well defined.

\section{Thick fakeons}

\label{thick}\setcounter{equation}{0}

In this section we generalize the identities to complex fakeon frequencies
and study the thick fakeons, which have nonvanishing widths at the tree
level. An example is offered by a propagator of the form%
\begin{equation}
\frac{2iM^{2}}{(p^{2}-\mu ^{2})^{2}+M^{4}}=\frac{1}{p^{2}-\mu ^{2}-iM^{2}}-%
\frac{1}{p^{2}-\mu ^{2}+iM^{2}},  \label{pthick}
\end{equation}%
where, as usual, $p^{\mu }=(e,\mathbf{p})$. We can view it as the difference 
$P_{\text{thick}}-\bar{P}_{\text{thick}}$, where 
\begin{equation}
P_{\text{thick}}=\frac{i}{2\Omega }\left( \frac{i}{e-\Omega }-\frac{i}{%
e+\Omega }\right)
\end{equation}%
and $\Omega =\sqrt{\mathbf{p}^{2}+\mu ^{2}-iM^{2}}$. The propagator $P_{%
\text{thick}}$ is of the type already studied, apart from two features: the
complex frequency $\Omega $ and the overall factor $i/(2\Omega )$.

More generally, the typical propagator of thick fakeons reads%
\begin{equation}
\frac{C}{p^{2}-\mu ^{2}-iM^{2}}-\frac{C^{\ast }}{p^{2}-\mu ^{2}+iM^{2}}%
=C^{\ast }P_{\text{thick}}-C\bar{P}_{\text{thick}},  \label{p2}
\end{equation}%
where $C$ is a complex number. Another interesting example is%
\begin{equation}
\frac{i(p^{2}-\mu ^{2})}{(p^{2}-\mu ^{2})^{2}+M^{4}}=\frac{i}{2}\left( \frac{%
1}{p^{2}-\mu ^{2}-iM^{2}}+\frac{1}{p^{2}-\mu ^{2}+iM^{2}}\right) =-\frac{i}{2%
}(P_{\text{thick}}+\bar{P}_{\text{thick}}).  \label{p3}
\end{equation}%
Although (\ref{p2}) and (\ref{p3}) cannot be obtained from a local
Lagrangian, they can be used to include certain limitations due to the
experimental apparatus, like the energy resolution around the
\textquotedblleft fakeon peak\textquotedblright .

We show that the spectral optical identities for diagrams involving
propagators like (\ref{pthick}), (\ref{p2}) and (\ref{p3}) can be derived as
(complex) linear combinations of (the complexified versions of) the
identities derived in the previous sections.

Typically, propagators like (\ref{pthick})\ appear in higher-derivative
theories. Examples are the Lee-Wick (LW) models \cite%
{leewick,lee,nakanishi,CLOP,grinstein}. It is worth to emphasize, though,
that the LW models do not have fakeons, but unstable ghosts, which are not
really out of the physical spectrum (in the same way as the muon is not out
of the physical spectrum of the standard model). The LW ghosts can be
dropped from the spectrum only in an effective field theory approach, if
they have relatively large decay widths.

Instead, fakeons are eradicated from the theory at the fundamental level
(and therefore, at all energies). Due to this, they do not need to decay.
For example, in the models of \cite{Tallinn1}, which have a $\mathbb{Z}_{2}$
symmetric fakeon sector, they have indentically vanishing widths. Moreover,
fakeons do not need higher derivatives, nor negative residues in front of
their propagators. What makes all this possible is the fakeon projection,
which is consistent only if we adopt the fakeon quantization prescription%
\footnote{%
In this paper, these two operations are condensed into a unique operation,
which is step 5) of subsection \ref{strat}.}.

As shown by Piva and the current author in \cite{LWformulation,LWunitarity},
it is possible to reformulate the Lee-Wick models, and a variety of other
higher-derivative theories, by converting them into theories of particles
and thick fakeons. This is one reason why it is interesting to generalize
the results of the previous sections to this type of fakeons. Other reasons
will be mentioned soon.

\bigskip

The generalization proceeds as follows. Consider a diagram $G$ involving
only real frequencies. Assume, for definiteness, that the legs 1 and 2 are
fakeons\footnote{%
We need two fakeons to highlight properties that are not visible with just
one. The generalization from two to an arbitrary number is then
straightforward.} and complete the strategy outlined in subsection \ref%
{strat}. Steps 1-4) give the spectral optical identities of the Feynman
version of $G$ (where all the internal legs are prescribed \`{a} la
Feynman). After that, we apply the fakeon prescription/projection of step 5).

Once these operations are concluded, we obtain the identities we start from
to make the complexification. We write them as%
\begin{equation}
f(\omega )+\bar{f}(\omega )+\sum_{c}f_{c}(\omega )=0,  \label{ide}
\end{equation}%
where $\omega $ is the frequency we want to complexify and $f(\omega )$, $%
\bar{f}(\omega )$ and $f_{c}(\omega )$ are the contributions of the diagram
itself, its complex conjugate and the cut diagrams, respectively. Note that,
after step 5), the thresholds do not involve fakeon frequencies any longer.
This is crucial, because it would make no sense to complexify frequencies
that appear inside delta functions.

We first consider the complexification $\omega \rightarrow \Omega $ of the
frequency $\omega $ associated with leg 1. Since (\ref{ide}) holds for
arbitrary $\omega $ (and $\omega $ does not enter the $\delta $ functions),
we can replace it everywhere by the complex $\Omega $. The principal values
may sound redundant, at this stage, but we keep them anyway, for reasons
that become clear below. Let us also multiply by a complex factor $K$. This
gives the relation%
\begin{equation}
Kf(\Omega )+K\bar{f}(\Omega )+\sum_{c}Kf_{c}(\Omega )=0  \label{idea}
\end{equation}%
with one caveat: the complex conjugation in $\bar{f}(\Omega )$ acts on the
function $f$, but not its variable $\Omega $. This means that (\ref{idea})
cannot be interpreted as the spectral optical identity of the diagram with
complexified frequencies, since $K\bar{f}(\Omega )$ is not the complex
conjugate of $Kf(\Omega )$. Indeed, a propagator with complex frequencies
cannot come from a Hermitian theory, so it cannot lead to an optical theorem.

A second identity can be obtained from (\ref{ide}) by replacing $\omega $
everywhere with $\bar{\Omega}$ and multiplying by $\bar{K}$:%
\begin{equation}
\bar{K}f(\bar{\Omega})+\bar{K}\bar{f}(\bar{\Omega})+\sum_{c}\bar{K}f_{c}(%
\bar{\Omega})=0.  \label{ideb}
\end{equation}%
Summing (\ref{idea}) and (\ref{ideb}), we get%
\begin{equation}
\left[ Kf(\Omega )+\bar{K}f(\bar{\Omega})\right] +\left[ \bar{K}\bar{f}(\bar{%
\Omega})+K\bar{f}(\Omega )\right] +\sum_{c}\left[ Kf_{c}(\Omega )+\bar{K}%
f_{c}(\bar{\Omega})\right] =0,  \label{idu}
\end{equation}%
which is the desired identity. We want to show that the first bracket is the
diagram with the propagator%
\begin{equation}
\frac{2iK\Omega }{e^{2}-\Omega ^{2}}+\frac{2i\bar{K}\bar{\Omega}}{e^{2}-\bar{%
\Omega}^{2}}  \label{proK}
\end{equation}%
in leg 1, the second bracket is the complex conjugate diagram and the sum
collects the contributions of the cut diagrams. We go through the steps 1-5)
of subsection \ref{strat} once again and generalize them to the case at
hand. When we do so, we understand that leg 2 has been treated in a similar
way. We distinguish the frequencies of the two fakeon legs by means of the
subscripts 1 and 2.

\bigskip

Let us start from step 2). Recalling how the residue theorem is applied to
get to (\ref{ide}), we recognize that the integral on the loop energies with
the propagator (\ref{proK}) must be performed along the LW integration path
(see \cite{grinstein} or \cite{LWformulation} for details), which is, by
definition, the path that picks the residues we need\footnote{%
Thick fakeons and LW decaying ghosts have the LW\ integration path in
common, although they differ in the rest. The reason is that, as far as the
integrals on the loop energies are concerned, the Feynman integration path,
the integration path following from the Wick rotation and the LW integration
path are all the same thing.}.

Moving to step 3), we need identities like (\ref{smartid}) to eliminate the
denominators $\tilde{D}_{\text{pseudo}}$. Those manipulations are legitimate
only if the denominators never vanish, which is not guaranteed by a
nonvanishing $M$. Indeed, we may find fractions like\footnote{%
This is the part of the argument where we need at least 2 fakeons.}%
\begin{equation}
\frac{1}{E-\Omega _{1}-\bar{\Omega}_{2}+i\epsilon },  \label{singu}
\end{equation}%
where $E$ is a linear combination of energies and possibly (real)
frequencies. The denominator of (\ref{singu}) is singular in extended
regions that may intersect the Minkowskian region (which is the real
subspace $P_{R}$ of the space $P_{\text{ext}}$ of the complexified external
momenta), as can be seen in the bubble diagram studied in ref. \cite%
{LWformulation} (see also \cite{fakeons} and \cite{nakanishi}). In
particular, this may happen when the legs 1 and 2 propagate the same
particle (same $\mu $ and same $M$). The $i\epsilon $ does not help and can
be ignored here.

The point is that, as shown in \cite{LWformulation}, the LW integration path
is an incomplete prescription. We need to complete it by deforming the
integration domain on the space components of the loop momenta. The domain
deformation generates Cauchy principal values inside the integrals \cite%
{LWformulation,fakeons}. Since the spectral optical identities concern the
integrands, they formally do not change.

The algebraic manipulations that remove $\tilde{D}_{\text{pseudo}}$ are
legitimate in the Euclidean region of $P_{\text{ext}}$, which extends to the
Minkowskian subregion below each threshold: there the denominators of the
integrands never vanish. The other regions are reached from the Euclidean
one either analytically (if we need to cross a physical threshold or a
fakeon threshold that does not fall on $P_{R}$) or by means of the domain
deformation (if we need to cross a fakeon threshold that falls on $P_{R}$).

\bigskip

Recapitulating, step 3) of subsection \ref{strat} is performed in the
Euclidean region. As far as step 4) is concerned, we move from the
Minkowskian subregion that lies below each threshold to the regions above
the thresholds, reaching them one by one, in all directions.

When the threshold involves fakeon frequencies and falls on $P_{R}$, we
proceed by means of the domain deformation, which incorporates also the
fakeon projection of step 5). When the threshold is physical (i.e., it
involves no fakeon frequencies), we apply the threshold decomposition as
before. Finally, when the threshold involves fakeon frequencies and\ does
not fall on $P_{R}$, it is not a true threshold (precisely because it is
located away from $P_{R}$), so it poses no obstacle.

Next, we apply the manipulations of point 4.b) and finally step 5).
Formally, the manipulations of step 4.b) remain the same, since the domain
deformation, as recalled above, generates principal values inside the
integrals on the space components of the loop momenta. To leave the previous
formulas the same, it is convenient to extend the meaning of the principal
value sign $\mathcal{P}$ to include the operations just described.

In the end, we get exactly (\ref{idu}). The procedure just outlined returns
the correct diagram, as well as its complex conjugate and the cut diagrams.
So doing, we have spectral optical identities for theories of particles and
fakeons of all types.

The coincidence splitting method is also required in the presence of thick
fakeons, to deal with powers of fractions like (\ref{singu}): since the
domain deformation may generate principal values, powers of those may appear
in involved diagrams.

\bigskip

If some situations it may be necessary to take into account the effects of
the experimental apparatus on fakeons, like the energy resolution $\Delta E$
of the processes mediated by them. We can achieve this goal by turning an
ordinary fakeon into a thick one, switching to a propagator of type (\ref{p3}%
) with $M=\Delta E$. Similarly, we can turn a thick fakeon into a thicker
one by increasing $M$. Since $\Delta E$ does not enter the Lagrangian, it
can have a different impact on different thresholds, which means that we can
have a different $\Delta E$ in each level of the threshold decomposition of
a diagram and each spectral optical identity (as long it is the same
throughout the identity). Assuming an \textquotedblleft
optimized\textquotedblright\ experimental situation, we can take a nonzero $%
\Delta E$ only in the thresholds that concern the experiment we are making.
For example, if we are studying the production of a $\mu ^{+}\mu ^{-}$ pair
and the theory includes a fakeon channel f$_{\text{c}}$ mediating that
production, we just need $\Delta E$ in the contributions interested by f$_{%
\text{c}}$. Anywhere else, we can keep $\Delta E=0$ and use the coincidence
splitting method, when needed. The spectral optical theorem continues to
hold at $\Delta E\neq 0$.

Note that it is nontrivial to be able to include the energy resolution
without violating the optical theorem. The interaction with the experimental
apparatus is not required to be unitary, since the system can no longer be
viewed as an isolated one.

\section{Massless fields and infrared divergences}

\label{massless}\setcounter{equation}{0}

In this section we explain how to handle the threshold decomposition and the
fakeon prescription in the presence of massless fields.

Massless fields are responsible for infrared divergences in scattering
amplitudes \cite{infred}. The divergences only appear on shell, when we
integrate on the space components of the loop momenta and the phase spaces.
To identify potential sources of divergences, we study the infrared
behaviors of the skeletons. We first discuss the issue with physical
particles only, then include fakeons. For definiteness, we assume that the
massless fields are photons. Collinear gluons and gravitons can be treated
the same way (with angular resolutions and other types of resolutions, in
addition to the energy resolutions).

The diagram $G$ and its conjugate $\bar{G}$ are well defined, since they are
off-shell, but the cut diagrams $G_{c}$ may not be separately well defined,
since the cuts put various particles on shell. The cancellation of the
infrared divergences in the sum%
\begin{equation}
\sum_{c}G_{c}=-G-\bar{G}
\end{equation}%
is due to the usual compensation between soft and virtual photons. We can
regulate each $G_{c}$ by equipping the photons with small fictitious masses $%
m$. Then we obtain the identity%
\begin{equation}
\sum_{c}G_{c}(m)=-G(m)-\bar{G}(m),  \label{gcm}
\end{equation}%
with self-evident notation.

The energy resolution $\Delta E_{\text{IR}}$ of the detectors can be taken
into account to make the limits $m\rightarrow 0$ well defined. If a cut
diagram $G_{c}(m)$ involves the integral on the phase space of a given
photon $\gamma $, with momentum $\mathbf{k}_{\gamma }$ and mass $m_{\gamma }$%
, we write it as $G_{c}^{<}(m)+G_{c}^{>}(m)$, where $G_{c}^{<}(m)$ is the
integral on $|\mathbf{k}_{\gamma }|\leqslant \Delta E_{\text{IR}}$ and $%
G_{c}^{>}(m)$ is the integral on $|\mathbf{k}_{\gamma }|\geqslant \Delta E_{%
\text{IR}}$. If the leg $\gamma $ is uncut in $G_{c}(m)$, we take $%
G_{c}^{<}(m)=$ $G_{c}(m)$ and $G_{c}^{>}(m)=0$. By construction, $%
G_{c}^{>}(m)$ has a well-defined limit $m_{\gamma }\rightarrow 0$. We obtain%
\begin{equation}
G(m)+\bar{G}(m)+\sum_{c}G_{c}^{>}(m)+\sum_{c}G_{c}^{<}(m)=0.  \label{gcrm}
\end{equation}

In the presence of massless fields, we cannot imagine the asymptotic state\
of a particle (such as the electron) as just made of the particle itself. We
have to think of it as made of the particle and a cloud of soft photons
radiated by it. Guided by this, we can attach $G_{c}^{<}(m)$ to some other $%
G_{c^{\prime }}^{<}(m)$ and reorganize the last sum of (\ref{gcrm}) into a
sum of terms with well-defined limits $m_{\gamma }\rightarrow 0$. We can
also view the last sum as a unique, convergent contribution. We repeat this
construction for all the photons. At the end, the limit where the photons
become massless gives the optical theorem in the presence of massless fields.

\bigskip

Now, consider the case of massless fields $\varphi _{0}$ in the presence of
fakeons. The fields $\varphi _{0}$ themselves must be physical (not
fakeons). Indeed, massless fakeons are excluded, because they violate
causality at arbitrarily large distances (instead of just microcausality,
see for example \cite{causalityQG}). As before, we equip each $\varphi _{0}$
with a small fictitious mass $m$. In the massless limit $m\rightarrow 0$,
some (multi)thresholds $\Delta _{i}$ that are distinct at $m\neq 0$ may tend
to the same threshold $\Delta _{0}$ for low $\varphi _{0}$ energies\footnote{%
For example, consider the second diagram of fig. \ref{2selfB} with $\varphi
_{0}$ in the vertical internal line and identical massive fields in the
other internal lines. We can view this case as the third diagram of fig. \ref%
{BoxD} with $m_{1}=m_{2}=m_{4}=m_{5}$, $m_{3}=0$ and $%
p_{3}=p_{1}+p_{4}=p_{2}+p_{5}$. Then it easy to check that $\Delta
^{135}\simeq \Delta ^{12}$ for low $\varphi _{0}$ energies ($\omega
_{3}\simeq 0$).}. This means that the $\Delta _{i}$ can interfere\ with one
another after the massless limit, even if they do not interfere before. In
section \ref{degenerate} we have shown that the limit of coinciding
thresholds is regular, because it allows us to define powers of principal
values and $\delta $ functions\footnote{%
There, we had powers of principal values and $\delta $ functions with the
same momenta. In the case of coinciding thresholds due to the massless limit 
$m\rightarrow 0$, the powers originate from different loops. The extra loop
integrals have further smearing effects.}.

We know that the starting uncut diagram $G$ (defined by the Feynman
prescription everywhere) does not have infrared divergences. This is enough
to guarantee that the every threshold $\Delta _{0}$ of its threshold
decomposition is separately convergent. The reason is that the potentially
divergent behaviors of different levels cannot compensate one another, for
example through identities such as (\ref{idprin}). Indeed, the infrared
divergences, like the ultraviolet ones, are due to power counting behaviors
of the integrands and do not have imaginary parts. Without imaginary parts,
they cannot propagate to lower or higher levels.

Thus, the compensations must occur among the contributions of the same
level. Ultimately, they occur within the same threshold $\Delta _{0}$,
because each threshold is independent.

What happens can be described as follows. We know that $\Delta _{0}$ can
come from different thresholds $\Delta _{i}$ of the $m\neq 0$ diagram $G(m)$%
. Each $\Delta _{i}$ can separately have infrared divergences in the limit $%
m\rightarrow 0$. However, $G$ is free of them, so the infrared divergences
due to $\Delta _{i}$ must cancel out in $\Delta _{0}$.

Moreover, if $\Delta _{0}$ contains a fakeon frequency, all the thresholds $%
\Delta _{i}$ it comes from contain fakeon frequencies: since the massless
field $\varphi _{0}$ is not a fakeon, it cannot change the nature (physical
vs fake) of the thresholds it enters. If two thresholds $\Delta _{i}$ tend
to coincide for low $\varphi _{0}$ energies in the $m\rightarrow 0$ limit,
they must either be both physical or both fake. Therefore, the cancellation
of infrared divergences survives the fakeon prescription/projection and the
limit $m\rightarrow 0$, as we wished to show. The optical theorem, in the
form explained above, continues to hold.

The cancellation need not occur if the massless field is a fakeon. The
argument just given cannot guarantee that all the $\Delta _{i}$ are treated
the same by the fakeon projection in that case.

\section{Conclusions}

\label{conclusions}\setcounter{equation}{0}

We have shown that the whole unitarity problem in quantum field theory can
be reduced to a set of algebraic identities, which do not require to
integrate on the space components of the loop momenta, or the phase spaces
in cut diagrams, and hold for each threshold separately, for diagrams with
arbitrary derivative vertices (as long as the Lagrangian of the theory is
Hermitian) and for propagators with arbitrary residues, masses and
frequencies.

The key ingredient is a proper threshold decomposition, since different
thresholds do not interfere with one another. First, we integrate on the
loop energies by means of the residue theorem, which is an algebraic
operation. Second, we ignore the integrals on the space components of the
loop momenta or phase spaces and work on the skeleton of the diagram. Third,
we eliminate the pseudothresholds, which are unphysical, because they
involve differences of frequencies. Fourth, we perform the threshold
decomposition, which provides a separate optical identity for each
(multi)threshold. Fifth, we drop the thresholds involving frequencies
associated with the legs that we want to quantize as fakeons.

The threshold decomposition must be done with due care. For example, it is
not known a priori how to extract the purely virtual content of a diagram,
which is defined by the procedure itself.

The gain in insight is important. The spectral optical identities hold for
thick as well as non thick fakeons, with arbitrary residues on the poles of
the propagators. We can treat massless physical fields and even introduce
certain details of the experimental apparatus, like the energy resolution
around the fakeon peak, without violating the identities and, ultimately,
unitarity. In the whole analysis we never need to make a single non
algebraic operation.

Summing the spectral optical identities associated with a loop diagram, we
derive the spectral optical theorem obeyed by its skeleton. As soon as we
resume the integrals on the space components of the loop momenta and the
phase spaces, we obtain the usual optical theorem for amplitudes.

We have given explicit examples to show how the threshold decomposition
works and how the spectral optical identities are derived. At one loop\ we
have analyzed the bubble, the triangle, the box, the pentagon and the
hexagon. At two loops we have studied the triangle with \textquotedblleft
diagonal\textquotedblright\ and the box with diagonal. Whole classes of
diagrams with arbitrarily many loops are included straightforwardly. The
calculations provide formulas for the loop integrals with fakeons and relate
them to the known formulas for the loop integrals with physical particles.
It is possible to use the results of this paper to implement the fakeon
prescription in softwares like FeynCalc, FormCalc, LoopTools and Package-X.

Finally, the strategy we have elaborated also provides a general proof of
the identities for arbitrary diagrams.

\vskip12truept \noindent {\large \textbf{Acknowledgments}}

\vskip 2truept

We are grateful to U. Aglietti, L. Marzola, M. Piva and M. Raidal for
helpful discussions and the Laboratory of High Energy and Computational
Physics of NICPB (National Institute of Chemical Physics and Biophysics),
Tallinn, Estonia, for hospitality during the first part of this work.

\end{document}